\documentclass{optica-article}

\journal{opticajournal} 

\articletype{Research Article}

\usepackage{lineno}
\usepackage{subcaption}
\usepackage{physics}

\begin{document}

\title{Non-transverse Electromagnetic fields in micro and nano fibers}

\author{U. Saglam,\authormark{1,*} D. D. Yavuz,\authormark{2,$\dag$}}

\address{\authormark{1}Department of Physics, 1150 University Avenue, University of Wisconsin - Madison, Madison, Wisconsin 53706, USA}

\email{\authormark{*}usaglam@wisc.edu} 
\email{\authormark{$\dag$}yavuz@wisc.edu} 


\begin{abstract*} 
 We present an analytical and numerical study of electromagnetic modes in micro and nano fibers (MNFs) where the electric and magnetic fields of the modes are not necessarily orthogonal to each other. We first investigate these modes for different fiber structures including photonic crystal fibers. We then discuss two specific applications of these modes: (1) Generation of hypothetical axions that are coupled to the electromagnetic fields through the dot product of electric and magnetic fields of a mode, $\vec{E} \cdot \vec{B}$, and (2) a new type of optical traps (optical tweezers) for chiral atoms with magneto-electric cross coupling, where the confining potential again is proportional to $\vec{E} \cdot \vec{B}$. 

\end{abstract*}

\section{Introduction}

In recent decades, the developments in fiber-optic communication paved the way for much progress in both the physics and applications of  sub-wavelength diameter wave-guiding structures~\cite{MNF-rev, MNF-China}. The guiding of light in these structures typically requires smooth, adiabatically-deformed spatial profiles for the refractive index~\cite{MNF-rev, how-thin}. The evanescent field from these structures can be utilized for shorter response times, better resolutions, and low power consumption sensors~\cite{MNF-rev, MNF-rev-sensor-status} as well as for trapping, guiding and reflecting of neutral atoms in the evanescent field of the guided light~\cite{atom-trap-1, atom-trap-2, atom-trap-3, atom-trap-4, atom-trap-5, atom-trap-6, atom-trap-7, atom-trap-8, atom-trap-9, atom-trap-10, atom-trap-11, atom-trap-12, atom-trap-13, atom-reflection}. In addition to trapping, the MNFs can also be utilized for the enhancement and measurement of the quadrupole transitions of atoms, for example alkali atoms cesium (Cs) and rubidium (Rb)~\cite{quadrupole-1, quadrupole-2, quadrupole-3, quadrupole-4, quadrupole-5}.

Sub-wavelength wave-guiding structures can be fabricated by tapering of hydrogen flame-heated quartz fiber~\cite{MNF-rev}, a method we have also utilized in some of our group's experiments~\cite{Gold2022, Karpel2019}. When analyzing the propagation of light in these structures, one needs to design fibers with propagating modes that are beyond the weakly-guiding limit where slowly varying approximation fails ~\cite{MNF-China, love-waveguide}. This leads to modes that are typically referred to as hybrid propagating modes, which are combinations of the more-usual transverse-electric (TE) and transverse-magnetic (TM) modes. These modes' electric and magnetic fields start to demonstrate a counter-intuitive non-transverse behavior and the angle between the electric and the magnetic fields, as well as the angles between the fields and the propagation direction, start to deviate from the well-known value of $\pi/2$. While this counter-intuitive behavior is not widely known, several groups discussed the existence of parallel electric and magnetic fields in propagating waves for different structures, some of which can be counter-propagating circularly polarized waves or waves in which the Poynting vector time averages to zero~\cite{Zaghloul1988, Uehara1989, Shimoda1990, Nishiyama2015, Mochizuki2021, chu-ohkawa-1, lee-comments, chu-responds, zaghloul-buckmaster-comment, chu-ohkawa-respond-2, mitchell-letter}. A wave with $E||B$ condition can be produced with three twisted-mode lasers~\cite{twisted-mode,twisted-mode-laser-1,twisted-mode-laser-2,twisted-mode-laser-3,twisted-mode-laser-4}, and was used to trap neutral sodium atoms in a Magneto-Optical Trap (MOT)~\cite{twisted-mode-trap}. In addition, such fields are prominent in the investigation of force free fields in plasmas~\cite{plasma}. 

A key feature of non-transverse modes is that because the angle between the electric and the magnetic fields is different from $\pi/2$, the dot product between these fields, $\vec{E}\cdot \vec{B}$, is non-zero. One fundamental application of the scalar field $\vec{E}\cdot \vec{B}$ is in laboratory search for axions and axion-like particles \cite{axion-3, strong-cp-1, strong-cp-2}. Were they to exist, axions would solve the strong CP problem of particle physics and they also form a compelling candidate for the dark matter in the universe.  The axions interact with the electromagnetic fields through $\vec{E}\cdot \vec{B}$ and this dot product can be used to generate axions in the lab~\cite{yavuz-axion-1, yavuz-axion-2, axion-2, axion-3}. In a recent paper from our group, we numerically explored a possible experimental setup where one can generate and detect axions via long MNF structures, and the rate of axion production highly depends on the $\vec{E}\cdot \vec{B}$ term~\cite{deniz_shay}.

This paper is organized as follows: we first summarize the analytical and numerical solutions for the hybrid modes in traditional step-index fibers, and demonstrate the behavior of $\vec{E}\cdot \vec{B}$ term within and in the vicinity of the fiber core. Second, we numerically explore the behavior of the $\vec{E}\cdot \vec{B}$ term in different structures such as a rectangular wave guide and a photonic crystal fiber. For these numerical investigations, we use the commercially available finite-difference software COMSOL~\cite{comsol-handbook}. Finally, we will discuss two specific applications of such non-transverse fiber modes: (1) generation of hypothetical axions that we mentioned in the previous paragraph, and (2) a new type of optical traps (optical tweezers) for chiral atoms with magneto-electric cross coupling, where the confining potential for the atoms is proportional to $\vec{E} \cdot \vec{B}$ \cite{atom-trap-5, atom-trap-8, sikes}.

\section{Analytical Calculations of Non-transverse fields in Microfibers}

The propagation of light in wave guides and optical fibers has been studied in depth by a number of authors. When the refractive index difference between the core and the cladding is small, $(n_{core}-n_{clad}=\Delta n\sim 0)$, then the fiber is in the weakly-guiding regime (Here, the quantities $n_{core}$ and $n_{clad}$ are the refractive indices of the core and the cladding, respectively).  In this regime, one can use the paraxial approximation for the propagating waves and the characteristic equation of the fiber is simplified to \cite{saleh_teich, pedrotti, katsunari-waveguide}:

\begin{equation}
    \frac{J_\nu'(u)}{u J_\nu (u)}=-\frac{K_\nu'(\omega)}{\omega K_\nu (\omega)}
\end{equation}

where $u= a \sqrt{k^2n_{core}^2-\beta^2}$ and $\omega=a \sqrt{\beta^2-k^2 n_{clad}^2}$ in a fiber of core diameter $a$. The quantity $\beta$ is the propagation constant and $k=2 \pi / \lambda$ is the magnitude of the wave-vector for the light. When this equation is solved for $\beta$, the solution for the desired mode is acquired. Electromagnetic fields under these approximations behave almost as transverse electromagnetic (TEM) waves, having near-zero longitudinal (z-component) of electric and magnetic fields. However, as one starts decreasing the diameter of the fiber, the weakly guiding approximation breaks down and more of the propagating modes acquire non-transverse behavior~\cite{saleh_teich, love-waveguide, katsunari-waveguide, MNF-China}. In order to find propagating modes in fibers with small core diameter, one needs to increase the difference of the indices of refraction between the core and the cladding $(\Delta n)$. In a physical context, the propagation of light in fibers with small radii demonstrate hybrid mode behavior~\cite{love-waveguide}. To find the hybrid modes, one needs to produce a more general characteristic equation from the surface boundary conditions $D^{\perp}_{clad}-D^{\perp}_{core}=\sigma_{induced}$ (perpendicular to the surface)  and  $E^{\parallel}_{clad}-E^{\parallel}_{core}=0$ (parallel to the surface)~\cite{saleh_teich, katsunari-waveguide, love-waveguide, MNF-China}. The derived characteristic equation is:

\begin{equation}
\begin{split}
        &\left[  \frac{J_\nu'(u)}{u J_\nu (u)}+\frac{K_\nu'(\omega)}{\omega K_\nu (\omega)} \right] \left[\frac{J_\nu'(u)}{u J_\nu (u)}+\left(\frac{n_{clad}}{n_{core}}\right)^2\frac{K_\nu'(\omega)}{\omega K_\nu (\omega)} \right]  \\
        &=\nu \left(\frac{1}{u^2}+\frac{1}{\omega^2} \right) \left[\frac{1}{u^2}+\left(\frac{n_{clad}}{n_{core}}\right)^2\frac{1}{\omega^2} \right] \quad .
\end{split}
\end{equation}

After this characteristic equation is solved, the Electric and Magnetic fields for the corresponding mode in the core region of a fiber can be found analytically and they are~\cite{katsunari-waveguide, love-waveguide}:

\begin{equation}
    \begin{split}
        E_r&=-i\beta \frac{a}{u}\left[\frac{1-s}{2}J_{\nu-1}\left( \frac{u}{a} r\right)-\frac{(1+s)}{2}J_{\nu+1}\left(\frac{u}{a}r\right)\right]cos(\nu \theta +\psi)\\
        E_\theta&=i\beta \frac{a}{u}\left[\frac{1-s}{2}J_{\nu-1}\left( \frac{u}{a} r\right)+\frac{(1+s)}{2}J_{\nu+1}\left(\frac{u}{a}r\right)\right]sin(\nu \theta +\psi)\\
        E_z&=J_\nu \left(\frac{u}{a} r\right)cos(\nu \theta +\psi)\\
        B_r&=-i\mu_0\omega_0 \epsilon_0 n_{core}^2 \frac{a}{u}\left[\frac{1-s_1}{2}J_{\nu-1}\left( \frac{u}{a} r\right)+\frac{(1+s_1)}{2}J_{\nu+1}\left(\frac{u}{a}r\right)\right]sin(\nu \theta +\psi)\\
        B_\theta&=-i\mu_0\omega_0 \epsilon_0 n_{core}^2 \frac{a}{u}\left[\frac{1-s_1}{2}J_{\nu-1}\left( \frac{u}{a} r\right)-\frac{(1+s_1)}{2}J_{\nu+1}\left(\frac{u}{a}r\right)\right]cos(\nu \theta +\psi)\\
        B_z&=-\frac{\beta s}{\omega_0 }J_\nu \left(\frac{u}{a} r\right)sin(\nu \theta +\psi)
    \end{split}
    \label{eq:eq3}
\end{equation}

where 

\begin{equation}
    s=\frac{\nu \left(\frac{1}{u^2}+\frac{1}{\omega^2}\right)}{\left[\frac{J'_{\nu}(u)}{uJ_{\nu}(u)}+\frac{K'_{\nu}(\omega)}{\omega K_{\nu}(\omega)}\right]},\qquad     s_1=\frac{\beta^2}{k^2n_{core}^2}s 
\end{equation}

and

\begin{equation}
    \omega_0=\frac{2 \pi c}{\lambda} \quad .
\end{equation}

\begin{figure}
    \centering
    \includegraphics[width=\linewidth]{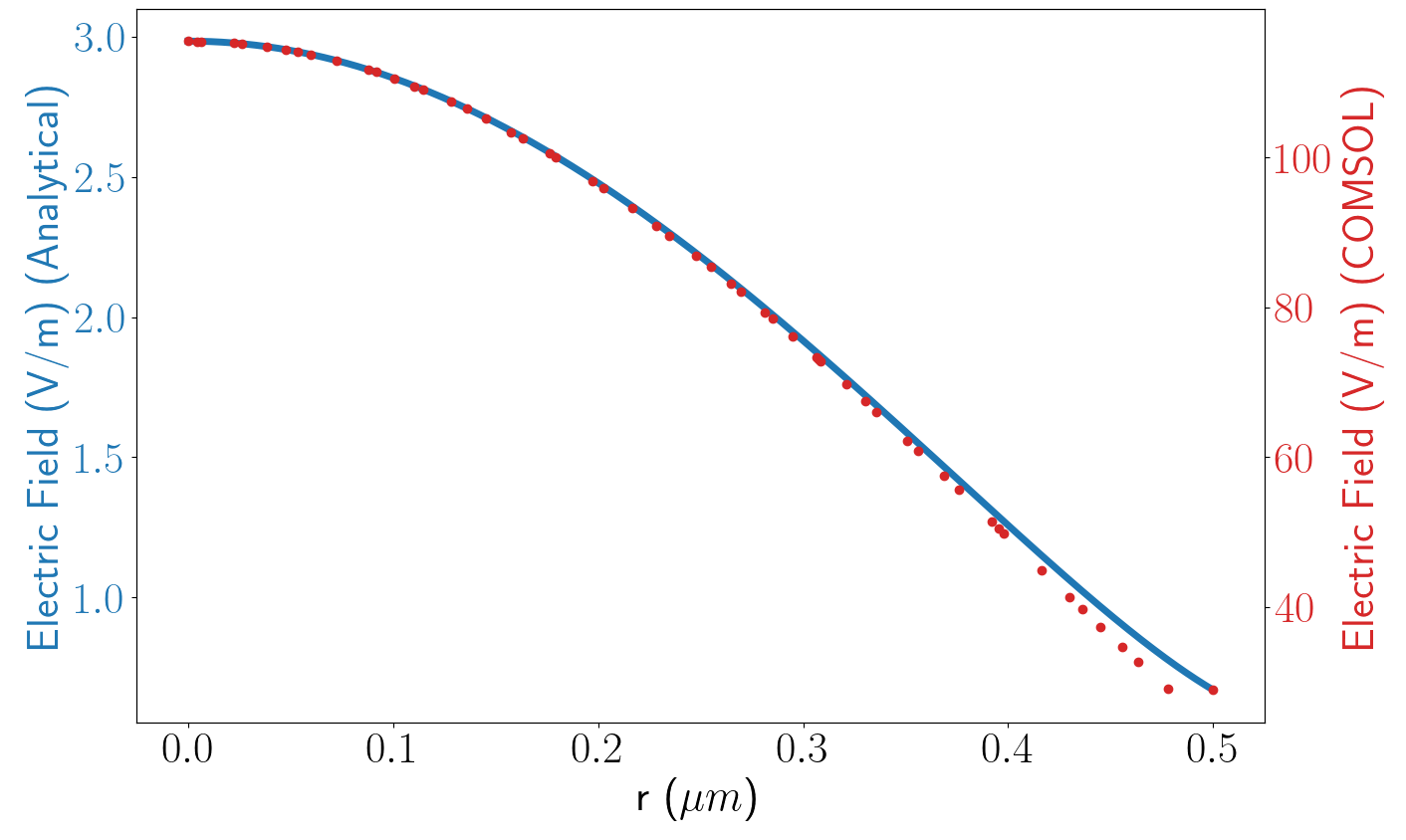}
    \caption{The plot for the Electrical field norm from normalized analytical calculations (solid blue line) and the same electric field data from COMSOL (dotted red line) with $a=0.5 \mu ms$, $\nu=1$ and $\beta=13.860\times10^6m^{-1}$. As expected, there is very good agreement between the analytical solution and the numerical calculation. }
    \label{fig:Enorm-fig1}
\end{figure}

As an example, we solve the full characteristic equation with $a=0.5 \mu ms$ along with $n_{clad}=1$ and $n_{core}=1.46$ with a wavelength of $\lambda=633$~nm, and calculate that the propagation constant for the lowest order hybrid mode $(HE_{11})$ is $\beta=13.860\times10^6 m^{-1}$. The analytically calculated radial profile for the Electric field of this hybrid mode is plotted in 
Fig.~\ref{fig:Enorm-fig1}. 

While it is possible to analytically calculate these modes for simple step-index structures, this is not the case for more complicated wave-guides including photonic crystal fibers. For much of what we discuss below, we will instead rely on numerically calculating these modes using commercial optical-wave simulation software, COMSOL. For comparison, in Fig.~\ref{fig:Enorm-fig1}, we also plot the profile of the electric field for the simple cylindrical step-index structure using COMSOL. As expected, there is very good agreement between the analytical solution and the numerical calculation.

In this paper, for various fiber structures that we discuss below, we will focus our attention to the lowest order mode (i.e., the fundamental mode) which is described thoroughly in Ref~\cite{MNF-China}. For structures that can maintain higher modes, the ratio of power carried by the evanescent field at the cladding and the field at the core can be modified to be higher or lower depending on the application~\cite{MNF-China, katsunari-waveguide, love-waveguide}.

Throughout this paper, we will summarize the non-transverse behavior of these waves propagating in the MNFs by examining the scalar field of 

\begin{equation} \label{eq:EdotB}
    F \equiv \text{cos}\theta=\frac{\vec{E} \cdot \vec{B}}{|\vec{E}||\vec{B}|} = \frac{E_x B_x^* + E_y B_y^* + E_z B_z^*+E_x^* B_x + E_y^* B_y + E_z^* B_z}{|\vec{E}||\vec{B}|} \quad . 
\end{equation}

Here, we define the quantity $\theta$, which is the geometric angle between the electric and magnetic field vectors. For simple step-index fiber structures in the weakly-guiding limit, we can use the solution given in above Eq.~(3) to analytically calculate this quantity:

\begin{equation}\label{eq:eq7}
    F=-\frac{\beta \epsilon_0 c  s\lambda}{4 \pi |\vec{E}||\vec{B}|}\left(J_\nu \left(\frac{u}{a}r\right)^2+J_{\nu-1} \left(\frac{u}{a}r\right)J_{\nu+1} \left(\frac{u}{a}r\right)\right)\text{Sin}( 2(\nu \theta + \psi)),
\end{equation}

where the norms of the fields, $|\vec{E}|$ and $|\vec{B}|$ can be written as, 

\begin{equation}
\begin{split}
|\vec{E} | |\vec{B}| &= (\mu_0/(4 \pi)) (((1/(
   u^2))(a^2 (-1 + s)^2 \beta^2 J_{\nu-1}((r u)/a)^2 + 
     a^2 (1 + s)^2 \beta^2 J_{\nu+1}((r u)/a)^2\\
     &+ 4 u^2 J_\nu ((r u)/a)^2 Cos(\nu \theta)^2\\
     &+2 a^2 (-1 + s^2) \beta^2 J_{\nu-1}( (r u)/a) J_{\nu+1}((r u)/a) Cos(2 \nu \theta)))) \\
     &((1/(u^2 \lambda^2 \mu_0))\epsilon_0 (a^2 (-4 n_{core}^2 \pi^2 \
J_{\nu-1}((r u)/a) \\
&+ (a (4 n_{core}^2 \pi^2 + 
            s \beta^2 \lambda^2) \nu J_\nu( (r u)/a))/(
         r u))^2 Cos(\nu \theta))^2\\
         &+ u^2 \beta^2 \lambda^4 J_\nu ( (r u)/
        a)^2 Sin(\nu \theta))^2 + 
      a^2 (s \beta^2 \lambda^2 J_{\nu-1}( (r u)/a)\\
      &- (a (4 n_{core}^2 \pi^2 + 
            s \beta^2 \lambda^2) \nu J_\nu ( (r u)/a))/(
         r u))^2 Sin(\nu \theta))^2)))^{1/2}
\end{split}
\end{equation}

\begin{figure}
     \centering
     \begin{subfigure}[b]{\textwidth}
         \centering
         \includegraphics[scale=0.2]{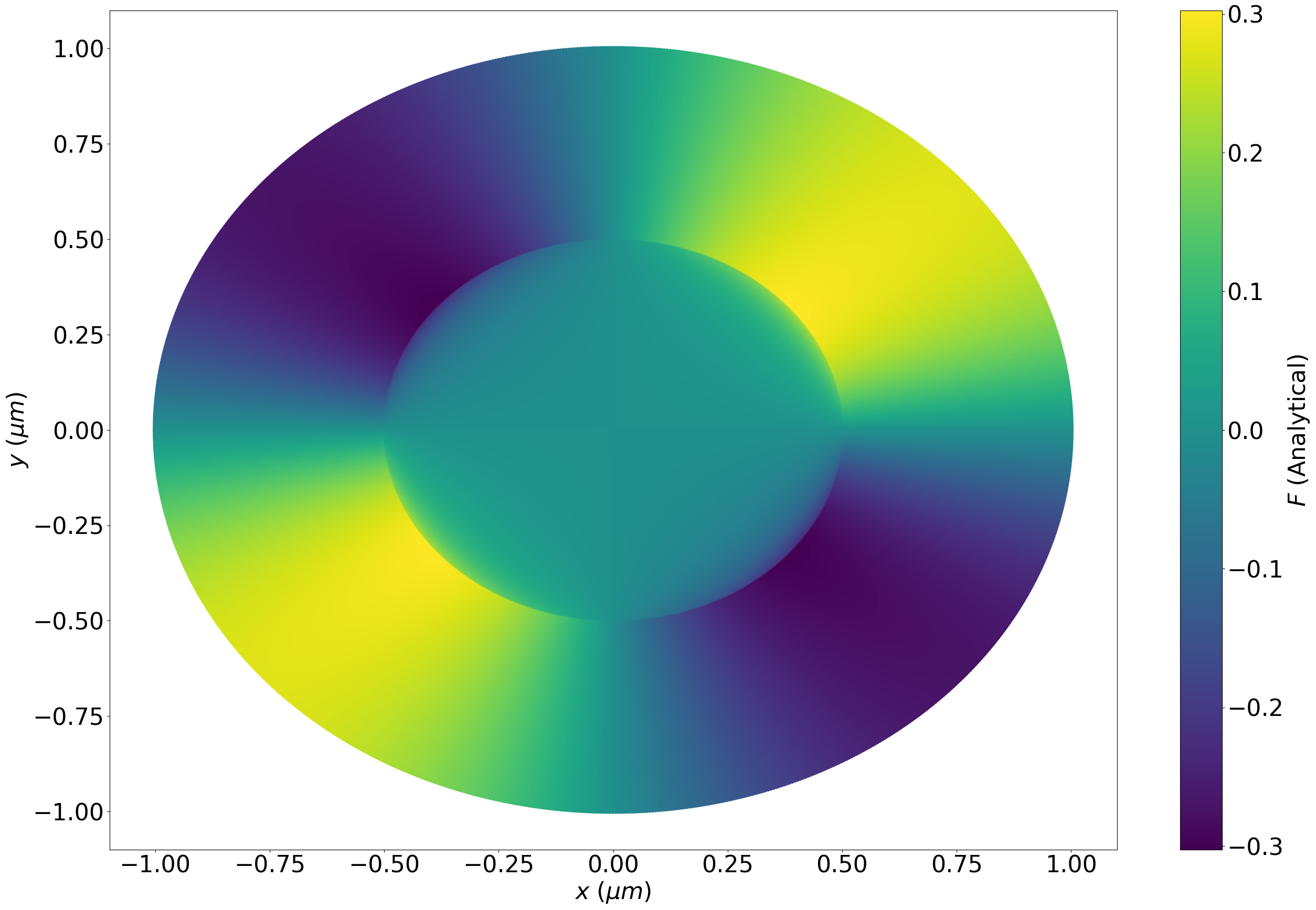}
         \caption{The scalar field $F$ from the analytical solutions of Eqs~(7) and (8). }
         \label{fig:EdotBScatter-analytical}
     \end{subfigure}
     \hfill
     \begin{subfigure}[b]{\textwidth}
         \centering
         \includegraphics[scale=0.2]{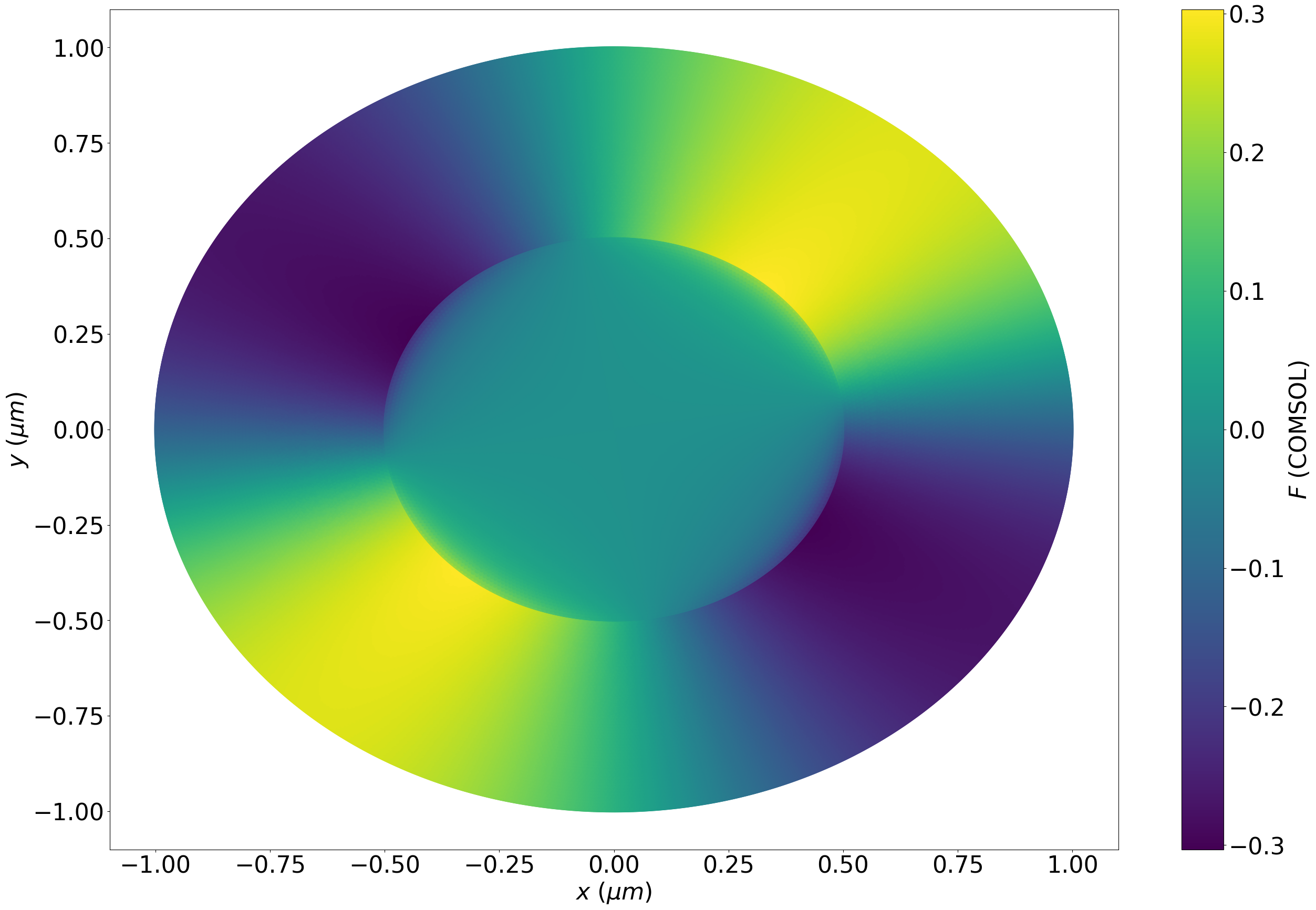}
         \caption{The scalar field $F$ from the numerical calculation using COMSOL.}
         \label{fig:EdotBScatter-comsol}
     \end{subfigure}
        \caption{False-color plot of the scalar field $F$ in the transverse $x-y$ plane within a region of $r=2a$ of the fiber from (a)  analytical solutions and (b) COMSOL calculations,  with $a=0.5~\mu ms$, $\nu=1$ and $\beta=13.860*10^6m^{-1}$. There is good agreement between the two results, both of which showing non-smooth behavior at the core-cladding boundary.  }
        \label{fig:three graphs}
\end{figure}

\begin{figure}
    \centering
    \includegraphics[scale=0.15]{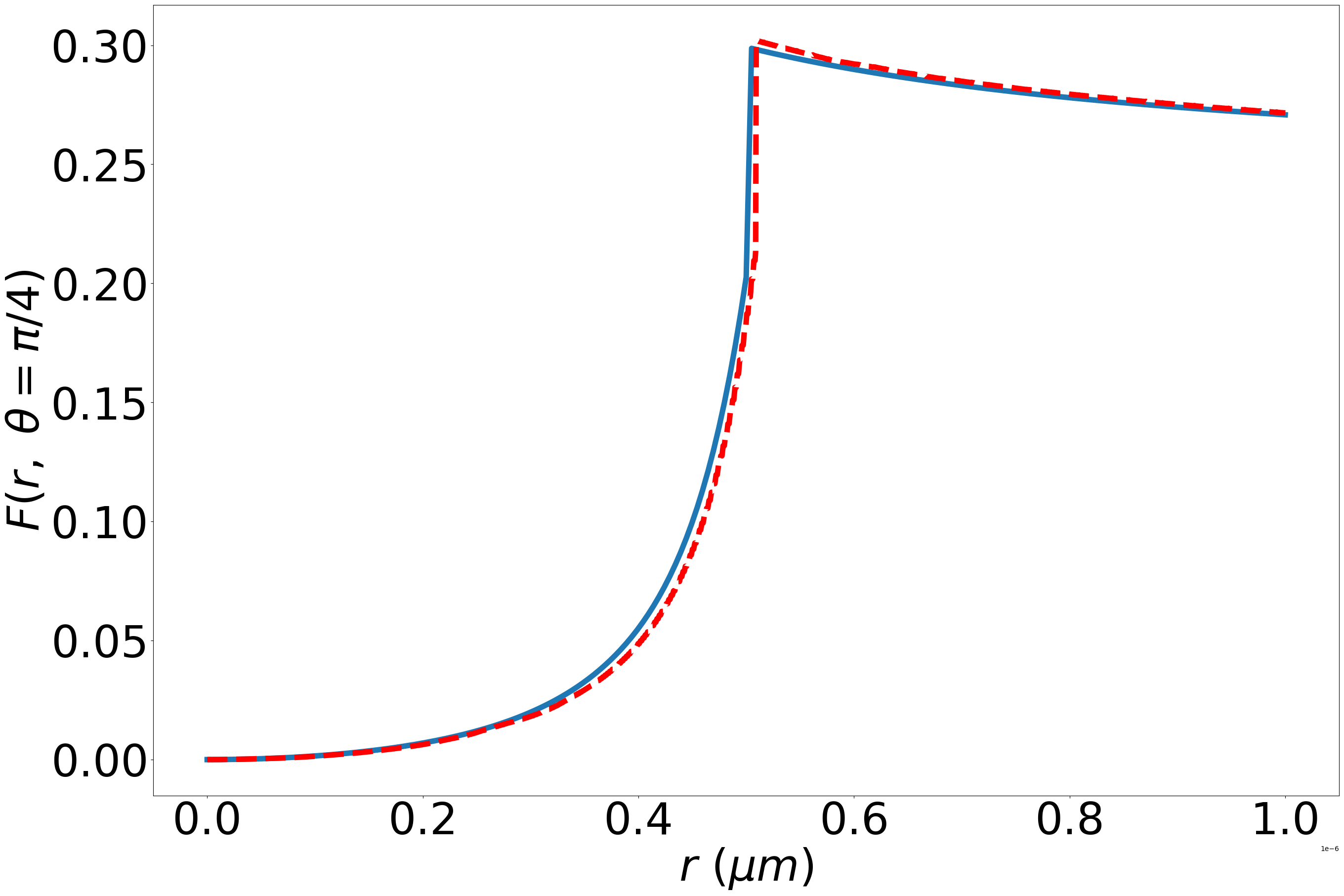}
    \caption{The plot for $F(r, \theta=\pi/4)$ with a=0.5 $\mu m$, $\nu=1$ and $\beta=13.860\times10^6m^{-1}$. The solid blue line is the analytical calculation and the dashed red line is the numerical results from COMSOL. The non-smooth behavior at the core-cladding boundary is due to the imposed Maxwell boundary conditions, where the perpendicular component of the displacement vector is continuous; but not the electric field vector. }
    \label{fig:EhPI4}
\end{figure}

We plot this scalar field from both analytical calculations (Fig.~\ref{fig:EdotBScatter-analytical}) and from COMSOL (Fig.~\ref{fig:EdotBScatter-comsol}) for the previous micro-fiber structure that we discussed in Fig.~1. These figures are false-color plots for $F$ in the transverse $x-y$ plane. As seen on these figures, there is again very good agreement between the analytical solution and the numerical calculation from COMSOL, both of which show nonzero values for the $F$ field. The electromagnetic fields start demonstrating non-transverse characteristics close to the core and cladding boundary. In Fig.~\ref{fig:EhPI4}, we plot the radial profile for the scalar field F at a specific azimuthal angle, $F(r, \theta)$ with $\theta=\pi/4$ for both COMSOL (dashed, red) and analytical (solid, blue) results. The results match well, both of which showing non-smooth behavior at the core-cladding boundary. The non-smooth behavior is due to the imposed Maxwell boundary conditions, where the perpendicular component of the displacement vector is continuous; but not the electric field vector. We observe that the discontinuity in the perpendicular component of the electric field becomes more drastic, as the refractive index difference between the core and the cladding gets larger~\cite{katsunari-waveguide}.

\begin{figure}[htbp]
    \centering
    \includegraphics[width=\linewidth]{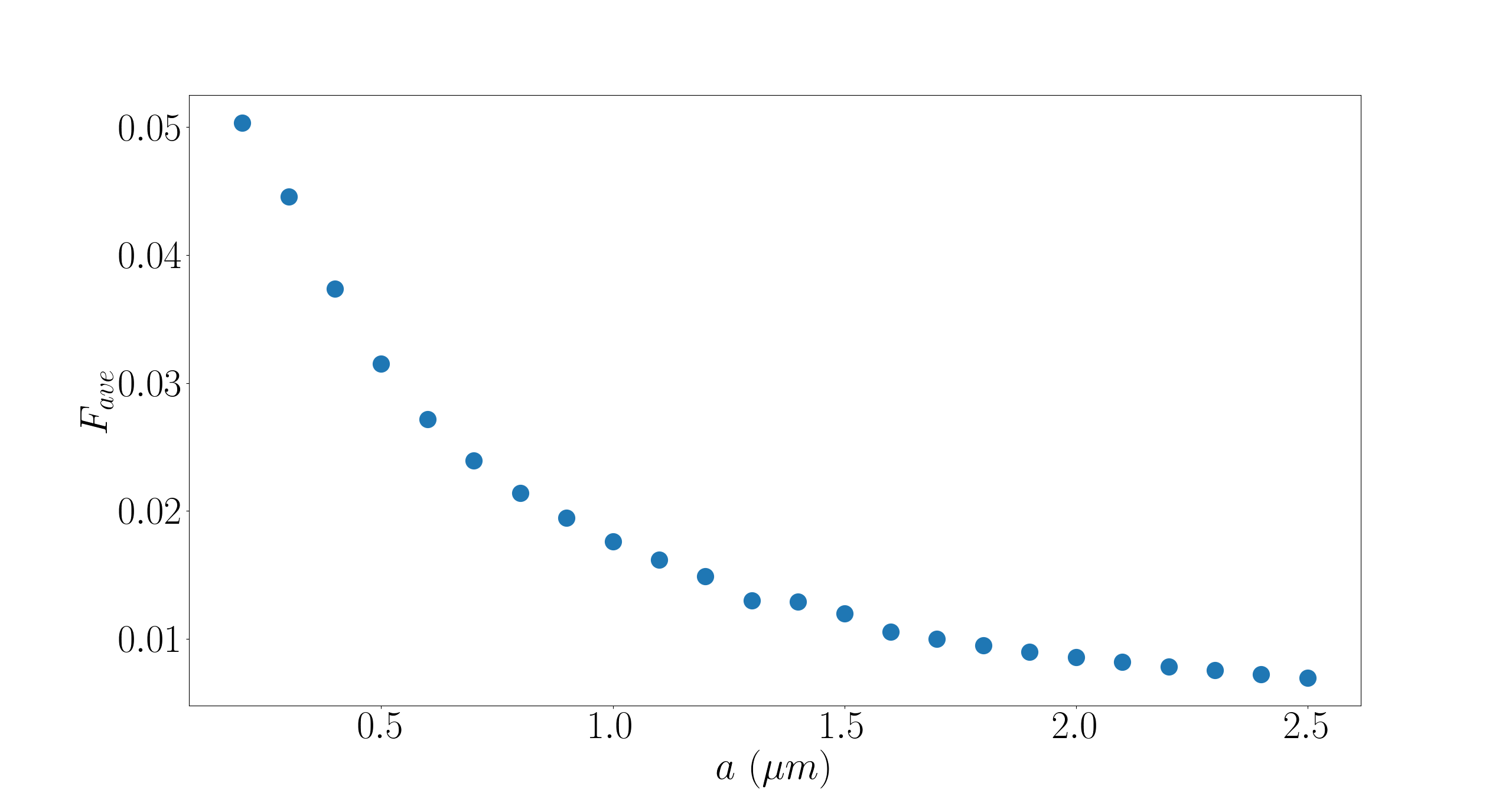}
    \caption{The plot of the average scalar field, $F_{ave}$ as a function of fiber core diameter $a$, for $\nu=1$ modes with $n_{clad}=1$ and $n_{core}=1.46$. The scalar field decreases as the core size increases, since waves approach transverse waves propagating in the bulk. }
    \label{fig:F_vs_rcore}
\end{figure}

A useful quantity that summarizes the non-transverse behavior of a fiber mode is the spatial average of the scalar field which we define as  $ F_{ave} \equiv \int dx dy \vert F(x,y) \vert /\int dx dy$. We expect this spatial average $F_{ave}$ to decrease as the core radius $(r_{core}=a)$ increases, since waves propagating in fibers with a larger core would approach transverse waves propagating in the bulk. To demonstrate this behavior, we increase $a$  and calculate the propagating modes in COMSOL. In Fig.~\ref{fig:F_vs_rcore}, we plot this quantity as a function of the core radius $a$, for the wavelength of light of $\lambda=1.55$~$\mu$m. We have verified that the critical parameter for this type of non-transverse behavior is the ratio $\lambda /a$ (i.e., the behavior at different wavelengths can be found by using an appropriate scaling of Fig.~\ref{fig:F_vs_rcore}). We find that, as expected, as the size of the core increases, the largely transverse waves near the central regions of the core results in an overall low value of the averaged scalar field F. We also numerically find that while one can find different modes with different $\nu$ and $\beta$ values,  the maximum value of the scalar field $F$ does not vary much between different modes (i.e. the behavior that is plotted in Fig.~\ref{fig:F_vs_rcore} is valid not only for the lowest order mode, but for other modes as well).

\begin{figure}[htbp]
    \centering
    \includegraphics[width=\linewidth]{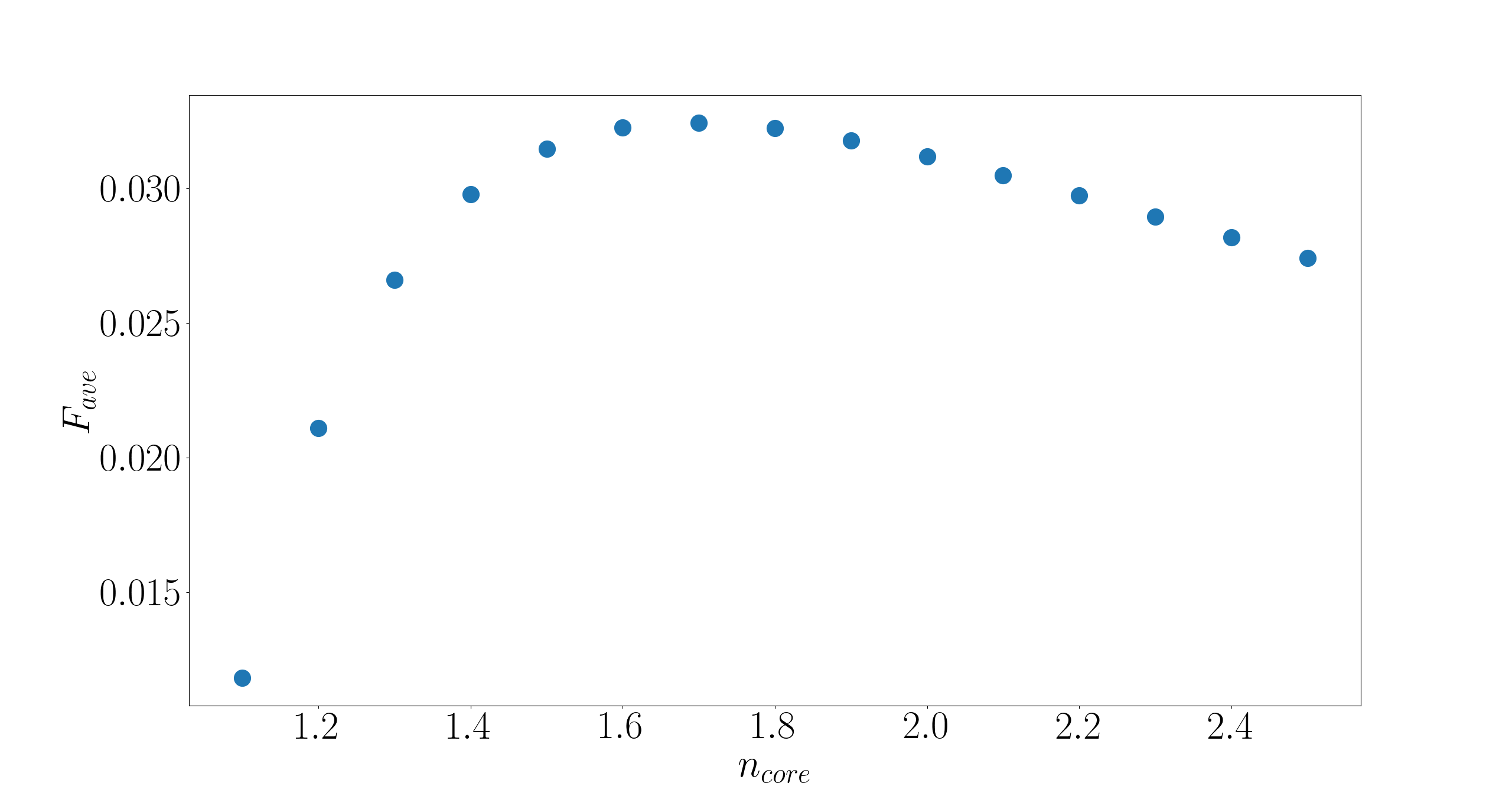}
    \caption{The plot of the average scalar field $F_{ave}$ as a function of the core refractive index, $n_{core}$, for $\nu=1$ modes with $r_{core}=0.5~\mu m$s (the refractive index of the cladding is fixed at $n_{cladding}=1$).}
    \label{fig:F_vs_nclad}
\end{figure}

We have also numerically investigated the behavior of the  scalar field $F$ with respect to the change in the refractive index of the cladding. In Fig.~\ref{fig:F_vs_nclad}, we plot the average scalar field, $F_{ave}$ as a function of $n_{clad}$, while keeping $r_{core}=a$ and $n_{core}=1.46$ constant. We observe that as $n_{core}$ increases the spatially averaged field F initially gets large and then later converges.  

\section{Different Structures for Non-transverse EM wave scheme}

In the previous section, we discussed non-transverse electromagnetic waves in a simple step-index fiber geometry where analytical solutions exist and we could, therefore, make comparisons with the numerical solutions using COMSOL. In this section, we will discuss an extension of these results to different wave-quide structures including rectangular and elliptical waveguides, as well as photonic-crystal (i.e. structured) fibers.

\subsection{Rectangular and Elliptical Waveguide}

Non-transverse EM waves in rectangular and elliptical waveguides may be useful in physical systems where an asymmetrical behavior between the two transverse axis is desired. In our numerical investigations of these systems, we have largely found the results that we have discussed in the previous section to be valid; i.e., we have observed similar behavior for these non-transverse waves as the size of the structure, or the refractive index difference between core and the cladding is varied. Because of this, we will not present a systematic study of these asymmetric structures. Rather, in this section, we will discuss the results for two specific asymmetric waveguides, as representative examples of what kind of non-transverse behavior can be expected. 
 
\begin{figure}
     \centering
     \begin{subfigure}[htbp]{\textwidth}
         \centering
         \includegraphics[scale=0.15]{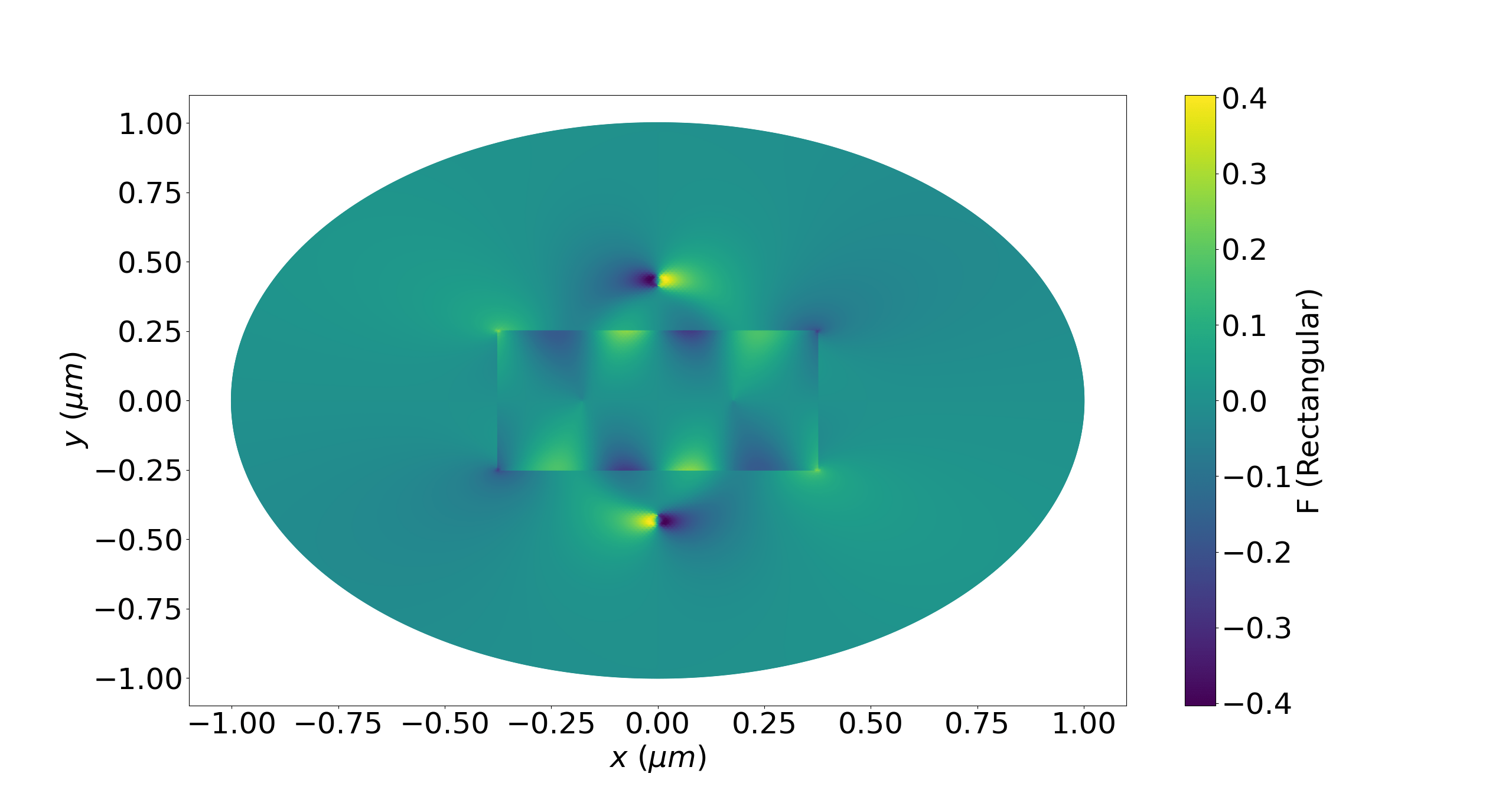}
         \caption{$F$ calculated by COMSOL with a propagation constant of $\beta=10.0035\times10^6m^{-1}$.}
         \label{fig:F_rect-a}
     \end{subfigure}
     \hfill
     \begin{subfigure}[htbp]{\textwidth}
         \centering
         \includegraphics[scale=0.15]{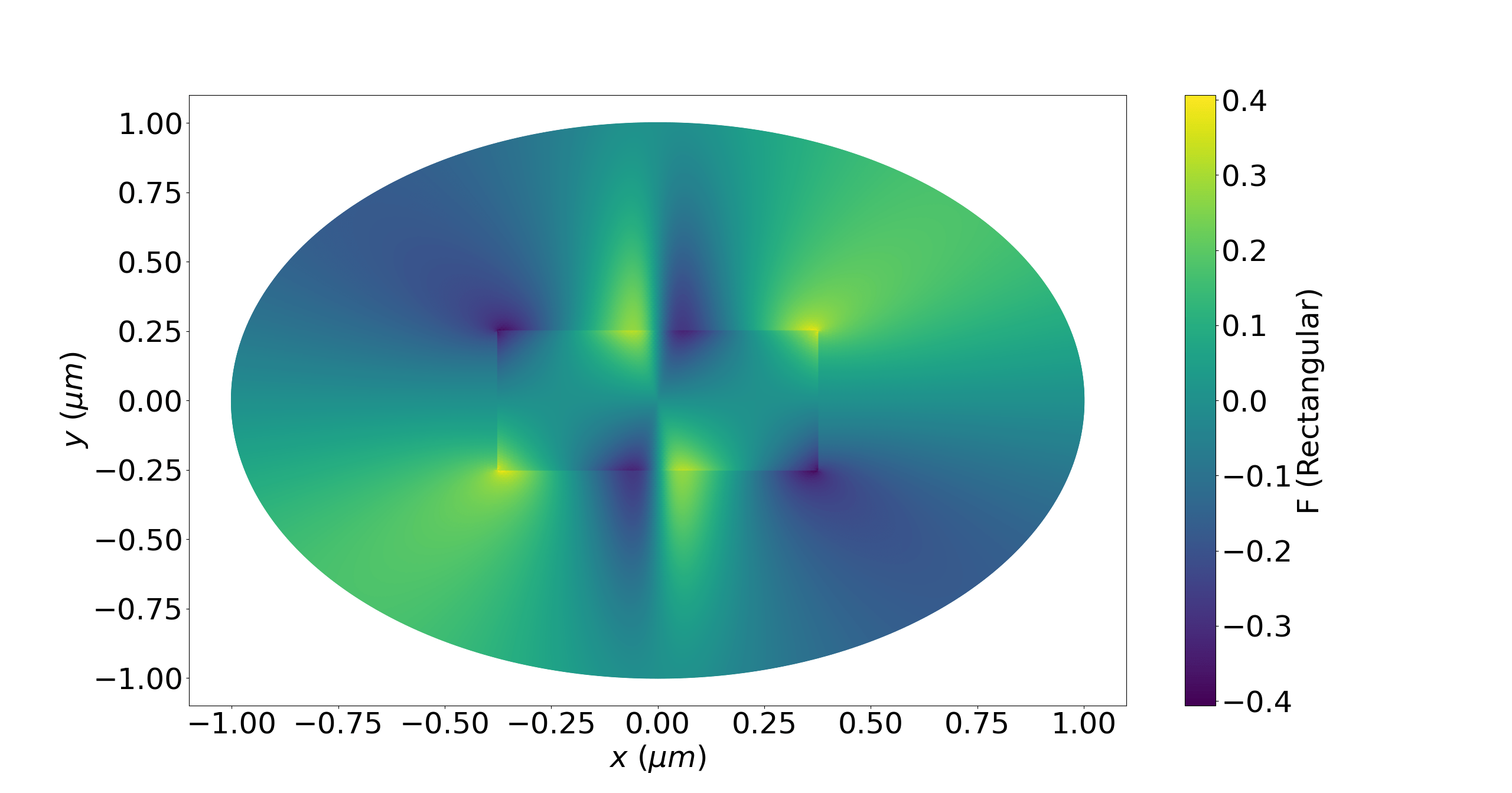}
         \caption{$F$ calculated by COMSOL with $\beta=11.7594\times10^6m^{-1}$.}
         \label{fig:F_rect-b}
     \end{subfigure}
      \begin{subfigure}[htbp]{\textwidth}
         \centering
         \includegraphics[scale=0.15]{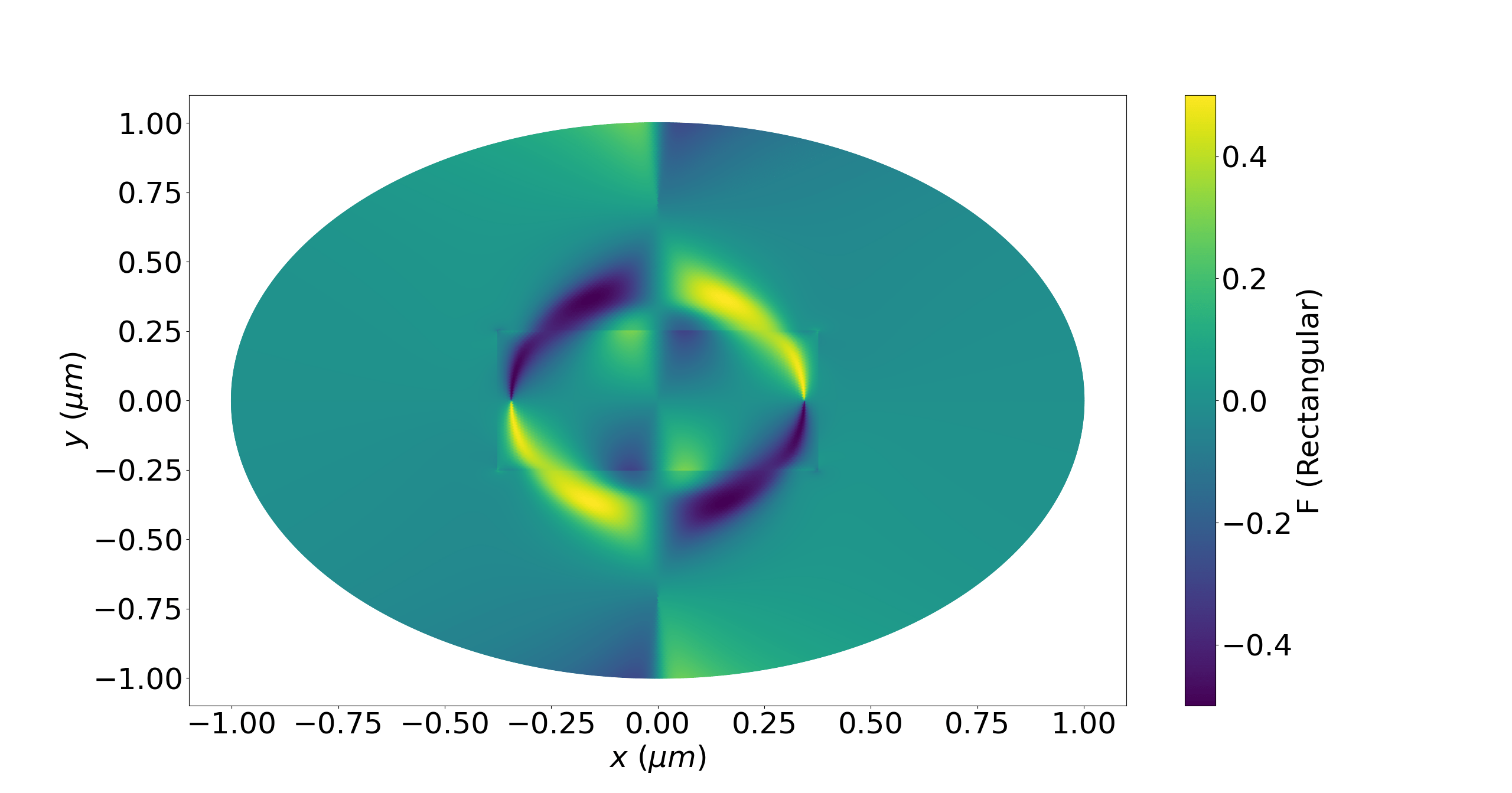}
         \caption{$F$ calculated by COMSOL with $\beta=9.80117\times10^6m^{-1}$.}
         \label{fig:F_rect-c}
     \end{subfigure}
        \caption{The false-color plot of $F$ given by Eq.~\ref{eq:EdotB} in a rectangular waveguide structure with $n_{core}=1.46$, $n_{clad}=1$, length$=0.75~\mu ms$, height$=0.5~\mu ms$, $\nu=1$ and $\beta$'s of $=10.0035\times10^6m^{-1}$(a), $11.7594\times10^6m^{-1}$(b), $9.80117\times10^6m^{-1}$(c)). }
        \label{fig:F_rect}
\end{figure}

 In Fig.~\ref{fig:F_rect}, we show results for a rectangular waveguide with a core length of $0.75 \mu ms$ and a core height of $0.5 \mu m$. Here, similar to the previous section we take the core and the cladding refractive indices to be $n_{core}=1.46$ and  $n_{cladding}=1$, and we plot the scalar field F for three different modes. These modes have propagation constants of $\beta$'s of $=9.80117\times10^6m^{-1}$(Fig.~\ref{fig:F_rect-a}), $9.80862\times10^6m^{-1}$(Fig.~\ref{fig:F_rect-b}), and $10.0035\times10^6m^{-1}$(Fig.~\ref{fig:F_rect-c}), respectively. In all of these figures, one can see the interesting spatial behavior of the field $F$, in addition to numerical values as high as 0.4.

In addition to a rectangular fiber, we have also investigate the behavior of $F$ in a fiber with an elliptical core with a major axis of $a=1\mu m$ and a minor axis of $b=0.5 \mu m$. The behavior of $F$ does not show a big difference compared to the circular-core fiber structure that we discussed above. We note that compared to the fibers with a circular core or the elliptical core, the rectangular core fiber displays a higher maximum value of $F$. This is because of the sudden change at the edges of the core.  

\begin{figure}[htbp]
    \centering
    \includegraphics[width=\linewidth]{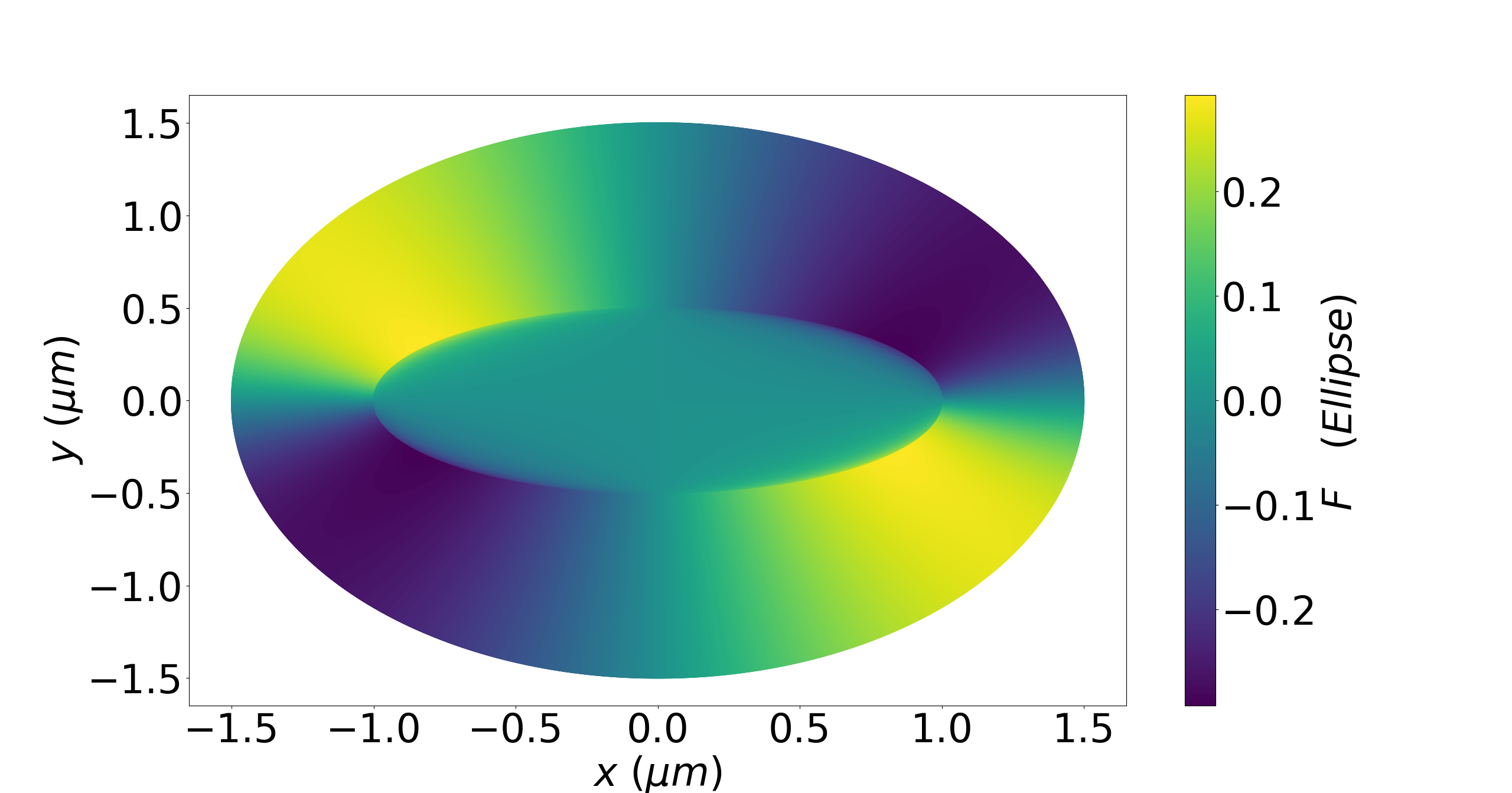}
    \caption{False-color plot of $F$ for a fiber with an ellipsoid core with  $\nu=1$, $n_{clad}=1$ and $n_{core}=1.46$ (the size of the major axis of the ellipse is $a=1\mu m$ and the minor axis size is $b=0.5 \mu m$. }
    \label{fig:F_ellipse}
\end{figure}

\subsection{Photonic Crystal Fibers}

Photonic crystal fibers are optical micro and nano sctructures where the index of refraction changes periodically~\cite{photonicCrystal-1, photonicCrystal-2}. Over the last two decades, the interest in these types of fibers has been continually growing, since their performance can substantially exceed traditional fibers. For example, one can manufacture photonic crystal fibers that can be single mode over a large wavelength range, or show highly-enhanced optical nonlinearities. 

\begin{figure}
     \centering
     \begin{subfigure}[htbp]{\textwidth}
         \centering
         \includegraphics[scale=0.15]{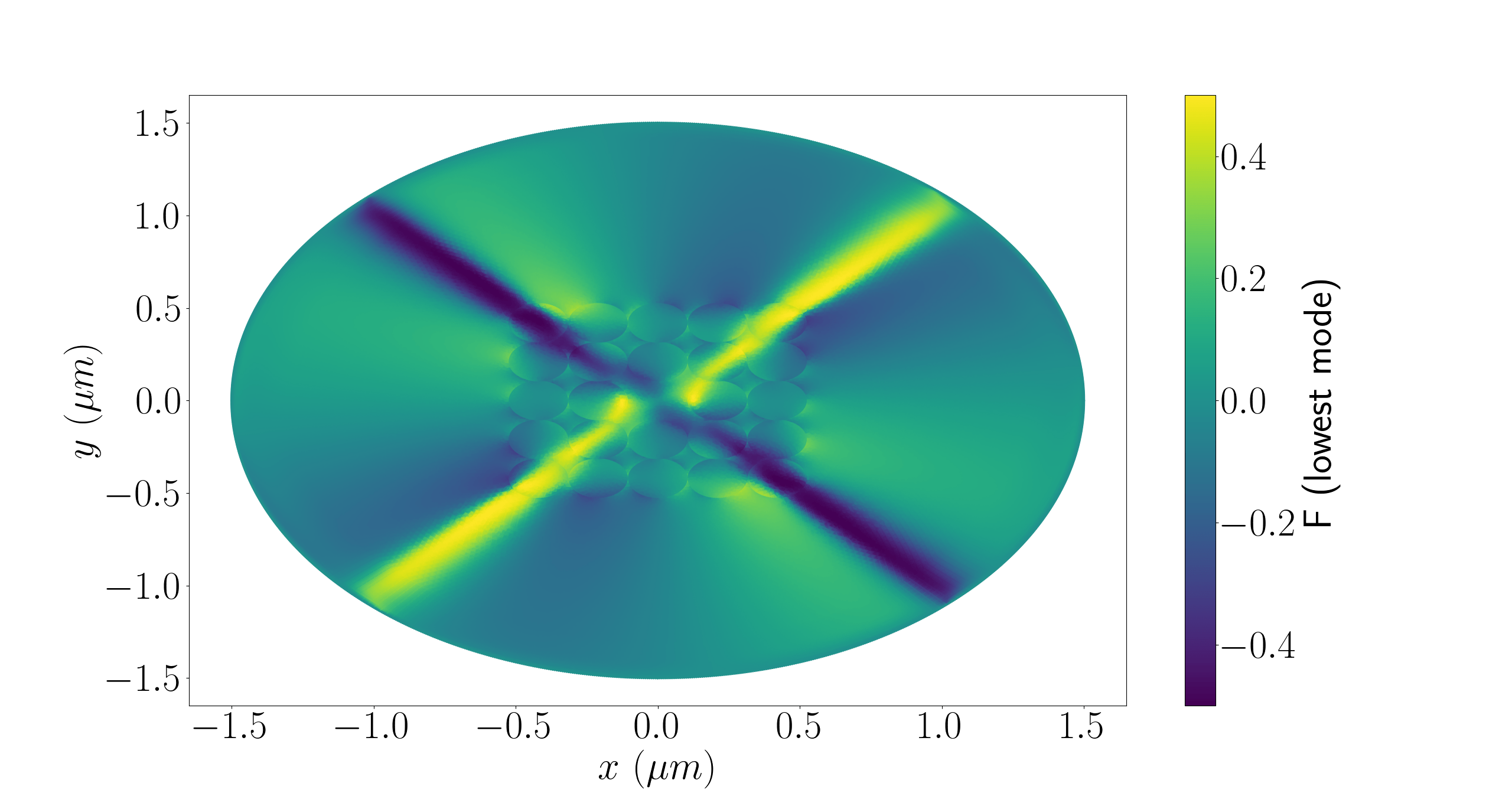}
         \caption{F of the lowest mode of a photonic crystal structure with $a=0.1 \mu m$  from the COMSOL data with $\beta= 10.5891\times10^6m^{-1}$.}
         \label{fig:F_photonic_crystal-a}
     \end{subfigure}
     \hfill
     \begin{subfigure}[htbp]{\textwidth}
         \centering
         \includegraphics[scale=0.15]{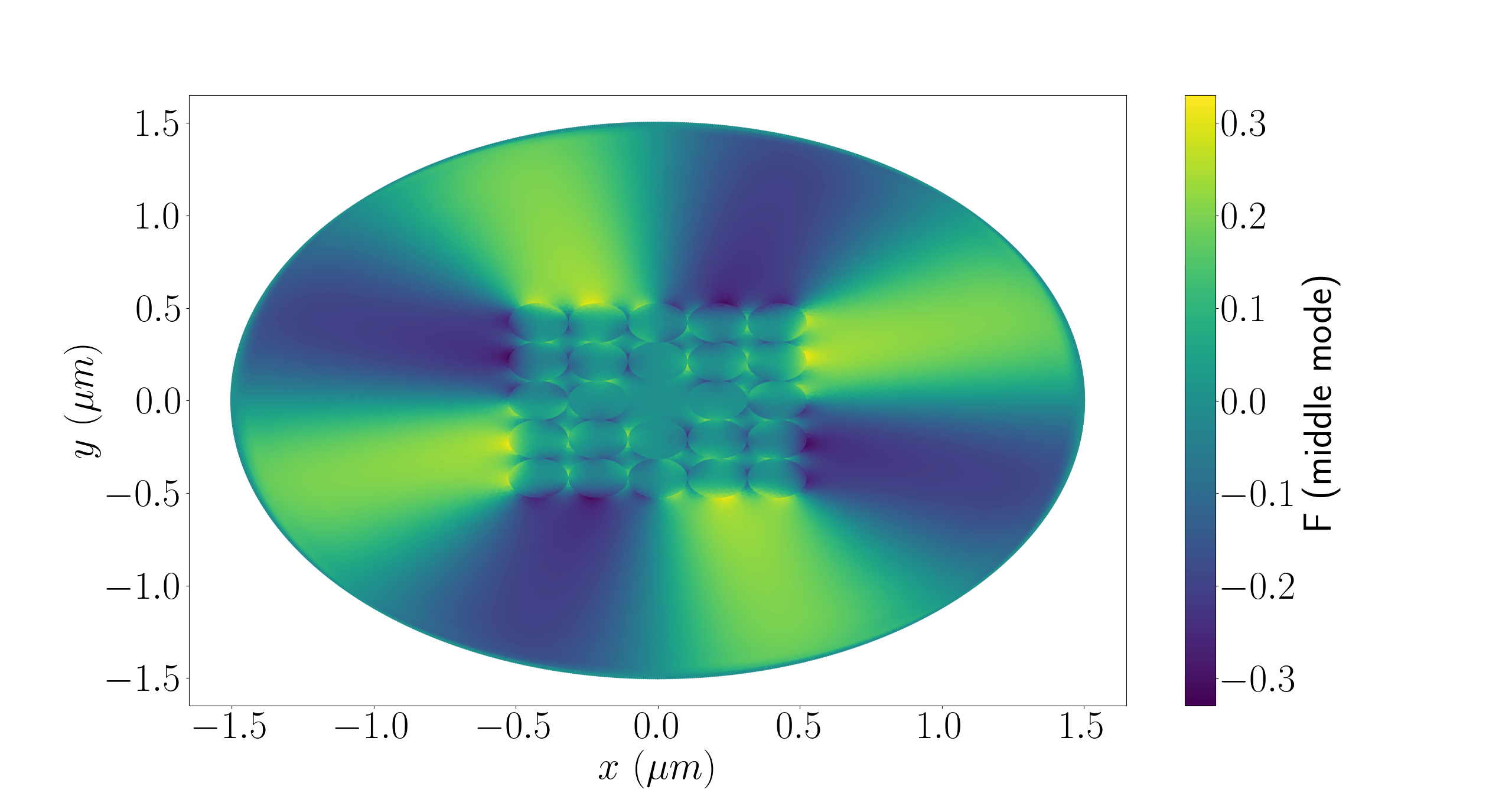}
         \caption{F of a middle mode of a photonic crystal structure with $a=0.1 \mu m$  from the COMSOL data with $\beta= 11.7207\times10^6m^{-1}$.}
         \label{fig:F_photonic_crystal-b}
     \end{subfigure}
      \begin{subfigure}[htbp]{\textwidth}
         \centering
         \includegraphics[scale=0.15]{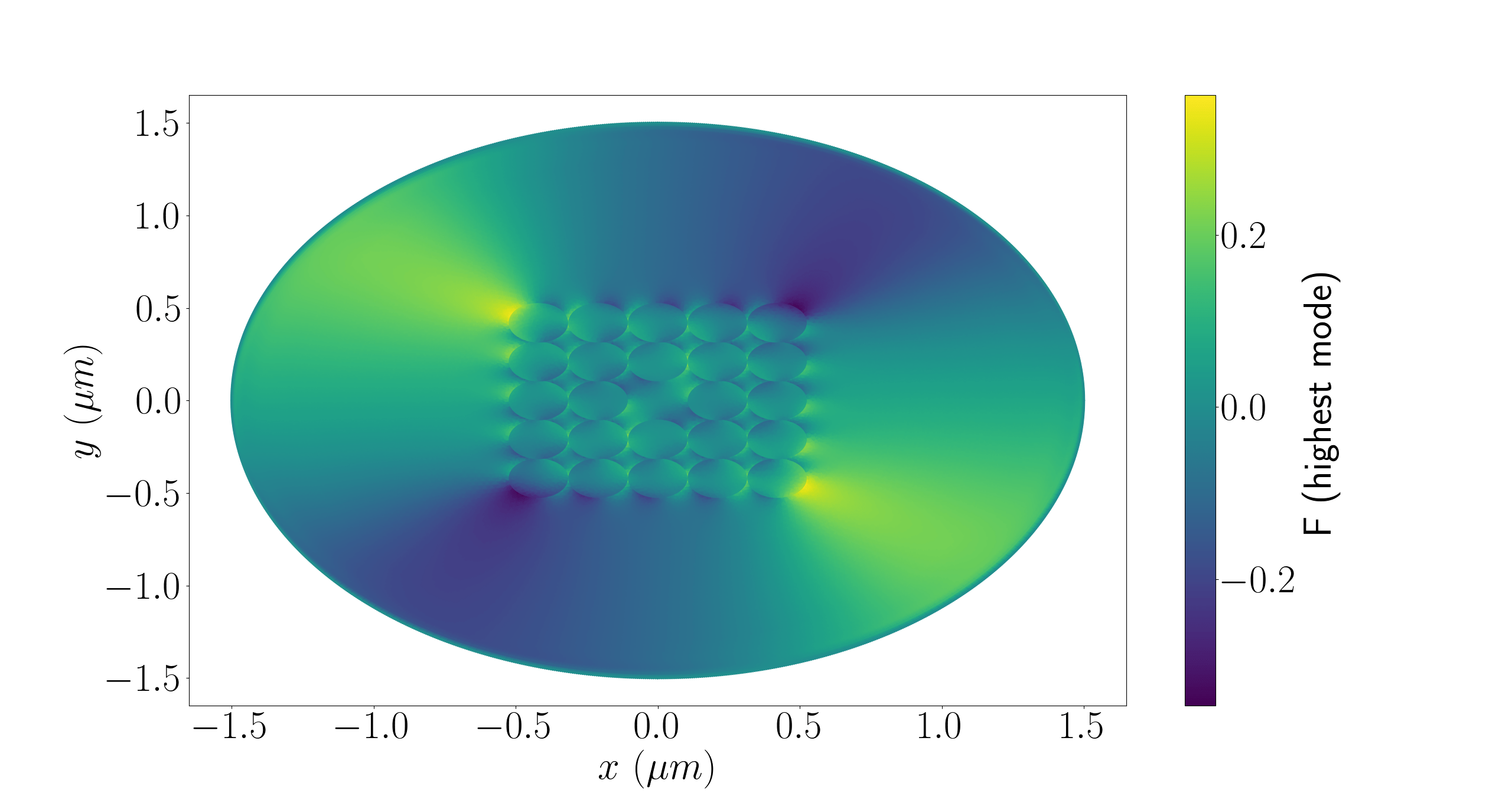}
         \caption{F of the highest mode of a photonic crystal structure with $a=0.1 \mu m$ from the COMSOL data with $\beta= 12.3381\times10^6m^{-1}$.}
         \label{fig:F_photonic_crystal-c}
     \end{subfigure}
        \caption{False-color plot of the scalar field $F$ calculated using Eq.~\ref{eq:EdotB} in a 5 by 5 photonic crystal structure with $n_{core}=1.46$, $n_{clad}=1$, $a=0.1~\mu ms$, $\nu=1$ and $\beta$'s of $=10.5891\times10^6m^{-1}$(a), $11.7207\times10^6m^{-1}$(b), $12.3381\times10^6m^{-1}$(c)). }
        \label{fig:F_photonic_crystal}
\end{figure}

In this section, we will discuss our COMSOL simulations that  investigate the behavior of $F$ in an example photonic crystal structure. We consider a photonic crystal fiber with an array of 5 by 5 cylinders. The diameter of each cylinder is $a=0.1 \mu m$ and the spacing between neighboring cylinders is $0.01 \mu m$. The cylinder at the center is missing since we aim to confine the modes near the centre of the structure. For a photonic crystal structure seen on Fig.~\ref{fig:F_photonic_crystal}, $\beta$'s of $=13.39381\times10^6m^{-1}$ (Fig.~\ref{fig:F_photonic_crystal-a}), $13.39411\times10^6m^{-1}$ (Fig.~\ref{fig:F_photonic_crystal-b}), $13.39471\times10^6m^{-1}$ (Fig.~\ref{fig:F_photonic_crystal-c}) modes' the scalar field $F$ has a maximum of about 0.3, also with reasonably high value for the $F$-field near the center of the structure. 

\begin{figure}[htbp]
    \centering
    \includegraphics[width=\linewidth]{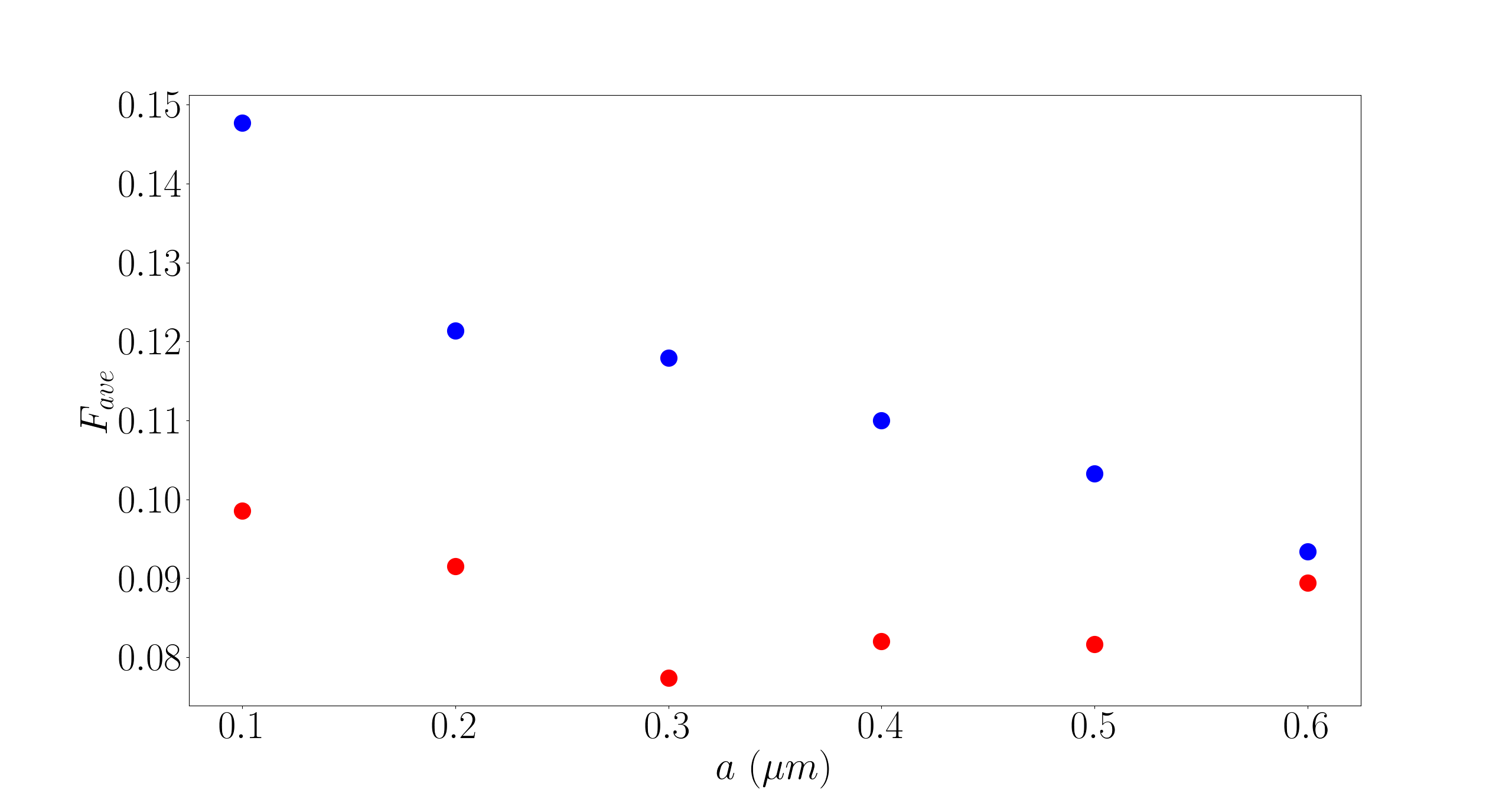}
    \caption{The plot of $F_{ave}$ averaged over a circular area of $R=5.1\times \sqrt{2} a$ vs a single fiber's $a$ for the 5 by 5 photonic crystal fiber structure. The blue dots are the results for the lowest mode while the red dots are the results for a representative middle mode between the highest and lowest order mode.}
    \label{fig:Fvsr_PC}
\end{figure}

We have also investigated the behavior of the average field $F_{ave}$, as we vary the radius of the core of the cylinders. In Fig.~\ref{fig:Fvsr_PC} we plot the quantity $F_{ave}$, which is the field $F$ is averaged over the area of a circle defined with a radius of $r_{av}=5.1 \times \sqrt{2}~a$. Here, we plot the results for the lowest mode (blue dots) and as well as a  representative middle-level mode (red dots), as we vary the core-radius of the cylinders, $a$. Here, we observe that the lowest mode has a higher average value of $F$. We also note that compared to Fig.~\ref{fig:F_vs_rcore}, the magnitude of $F_{ave}$ is three times larger for the lowest order modes, while the observed decay behavior is similar. We find that, while performing these calculations in COMSOL, as the cylinder radius $a$ gets larger, the simulation space gets large and rapidly increases the memory requirement of the computation. As an example, a photonic crystal structure with a single cylinder radius $a=0.5$ with a coarse mesh, most of the physical 32 GB RAM and around 80 GB of virtual RAM was used and the calculation took around 8 to 10 hours. Since we were working with distances that are comparable to the wavelength of the light, EW frequency domain (EWFD) analysis was required in COMSOL, which takes more time than the beam envelopes (EWBE) analysis~\cite{comsol-handbook}.

\begin{figure}[htbp]
    \centering
    \includegraphics[width=\linewidth]{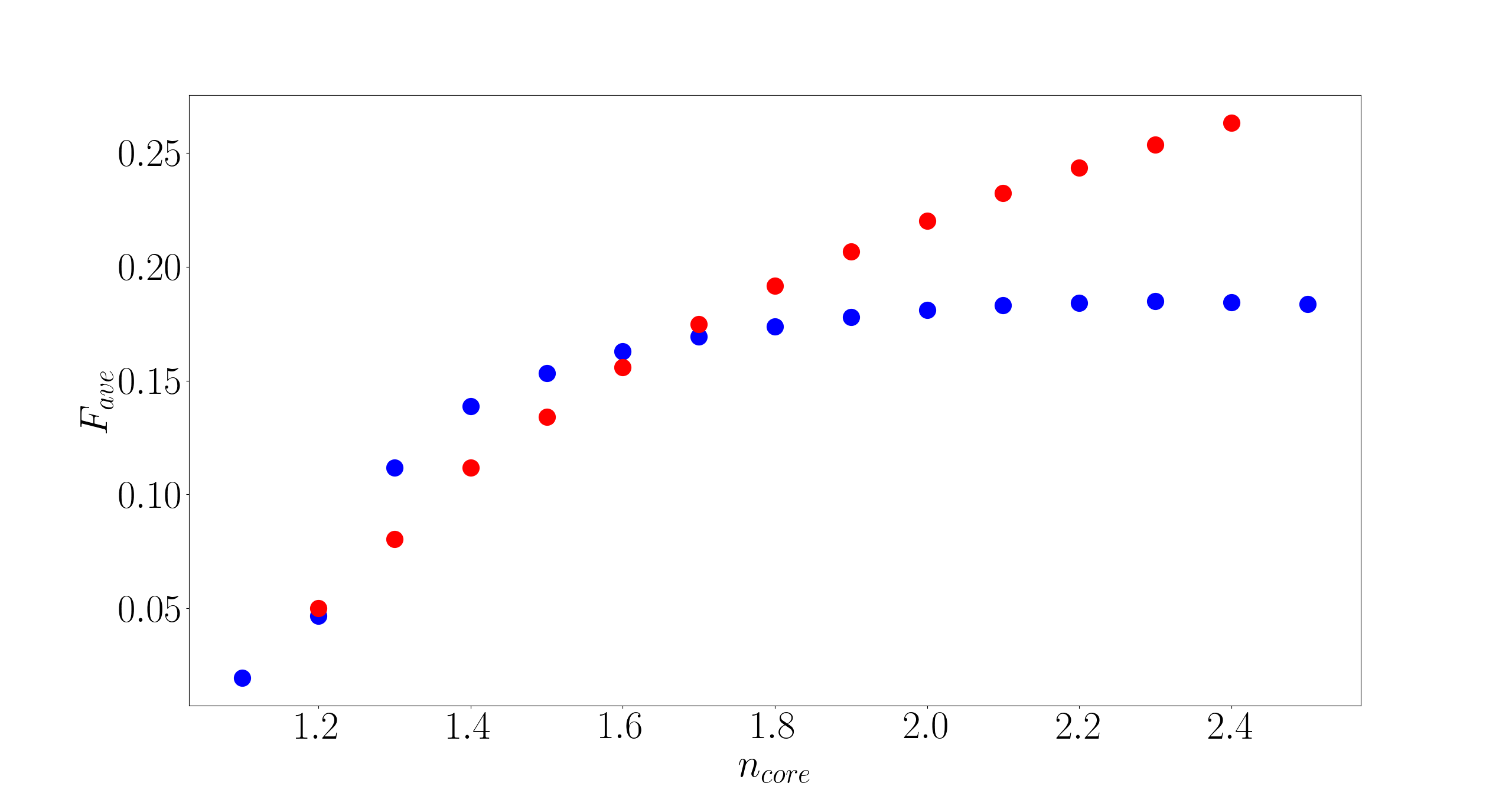}
    \caption{The plot of $F_{ave}$ averaged over a circular area of $R=5.1\times \sqrt{2} a$ vs the index of refraction of the cores $n_{core}$ for the 5 by 5 photonic crystal fiber structure. The blue dots are the results for the lowest mode while the red dots are the results for a representative middle mode between the highest and lowest order mode.}
    \label{fig:Fvsn_PC}
\end{figure}

To see the behavior of $F_{ave}$ with respect to $n_{core}$, we calculate the behavior of $F_{ave}$ in Fig.~\ref{fig:Fvsn_PC} with core radius $a=0.1~\mu m$, where blue dots being the lowest order mode and the red dots being a middle mode. Again, here, $F_{ave}$ is larger than a single-cored structure seen on Fig.~\ref{fig:F_vs_nclad}, while the overall behaviour of the dependence to  $n_{core}$ is similar.

\section{Generation and Detection of Hypothetical Axion Particles}

Since their first prediction about four decades ago, the interest in hypothetical axions has been continually growing \cite{review,amoreview,casper,irastorza}. Light axions or axion-like-particles with a mass in the $10^{-6} $~eV $< m<$ $10^{-2} $~eV range form a compelling candidate for the dark matter in the universe  \cite{darkmatter1,darkmatter2,darkmatter3}. The existence of axions would also solve one of the longstanding theoretical problems in the standard model of particle physics; the so-called strong CP problem \cite{axion-3,strong-cp-1,strong-cp-2}. Not surprisingly, there has been a large number of experimental efforts to detect this elusive particle.  The theoretical proposals exploring axion-photon coupling date back to 80's \cite{sikivie,krauss,maiani}, spurring much experimental work. One set of experiments aim to detect axions that are naturally present in the environment \cite{admx1,admx2,cast1,iaxo}. Another set of experiments work towards generating and detecting axions in the lab, and have greater control of their experimental parameters since they do not rely on an external source of axions   \cite{lsw1,lsw2}. This set of experiments are cordially referred to as {\it light shining through a wall (LSW)}, and their sensitivity has been steadily increasing over the last few decades.

As we will discuss below in detail, were they to exist, the axions are coupled to the fields of electromagnetics through the $\vec{E}  \cdot \vec{B}$ term. In LSW experiments, the generation of axions is typically accomplished by utilizing the interaction of the electric-field of a laser beam with an intense DC magnetic field. As we have noted in a recent publication, such generation could also be accomplished using both the electric field and magnetic field of a laser \cite{deniz_shay}. This requires a non-vanishing $\vec{E}  \cdot \vec{B}$ term due to a laser beam, which requires non-transverse electromagnetic modes. One application of the modes that we have discussed above is, therefore, to the generation of such hypothetical axions. 

 More specifically, the Klein-Gordon equation for the axion field  $\Phi (\vec{r},t)$  has a driving term that involves the dot product of the electric and magnetic fields:
\begin{eqnarray}
 \nabla^2 \Phi - \frac{1}{c^2} \frac{ \partial^2 \Phi}{\partial t^2}- \left( \frac{m c} {\hbar} \right)^2  \Phi =  \frac{g_{a \gamma \gamma}} {\mu_0 c} \vec{E} \cdot \vec{B} \quad .
\end{eqnarray}

\noindent Here, $m$ is the mass of the hypothetical axion and $\nabla^2$ is the Laplacian operator.  The quantity $g_{a \gamma \gamma}$ is the axion-EM coupling constant. As we discussed in our recent paper, the driving $\vec{E}  \cdot \vec{B}$ \cite{deniz_shay} term can be used to confine the generated axion wave, $\Phi$. Using procedures quite similar to finding the optical modes of a fiber \cite{saleh_teich}, the Klein-Gordon equation for the axion field can be reduced to a single radial differential equation, driven by the spatial mode profile of $\vec{E}  \cdot \vec{B}$. This radial differential equation can then be numerically integrated to find the shapes of the confined axion modes. 

In our recent paper, we assumed the simplest case of only radial dependence of the driving term and therefore found the cylindrically-symmetric solutions for the confined axions. For the driving laser beam in a hybrid mode, as shown in Eq.~(3), there is a dependence of the electric and magnetic fields, and therefore $\vec{E}  \cdot \vec{B}$ term, on the azimuthal angle $\theta$.

From Eq.~(\ref{eq:eq7}), since we know the general form of the driving term, we define our trial solution  $ \Phi(r, \theta, z)$ to be of the form:

\begin{equation}
    \Phi(r, \theta, z)= u_\phi(r, \theta)
\end{equation}

Differing from our recent paper where we considered excitation using two different laser beams \cite{deniz_shay}, this solution does not include a propagation term that is dependent on $z$.  This is because both the electric and magnetic fields have the same $z$-dependent propagation term, which cancels out in the dot product. Consequently, the Klein-Gordon equation of Eq.~(9) can be reduced to the following differential equation for $u_\phi(r, \theta)$:

\begin{equation}
    \frac{d^2 u_\phi}{dr^2}+\frac{1}{r}\frac{du_\phi}{dr}-\frac{1}{r^2} \frac{d^2u_\phi}{d\theta^2}+\left(\frac{mc}{\hbar}\right)^2  u_\phi = \frac{g_{a\gamma\gamma}}{\mu_0 c} E(r,\theta) B^*(r, \theta)
\end{equation}

 Next, we define the quantity, $\Delta k ^2 \equiv (mc/\hbar)^2$.  With this definition and using the analytical result from Eq.~(\ref{eq:eq7}) for $E \cdot B$, we obtain:

\begin{equation}
    \frac{d^2 u_\phi}{dr^2}+\frac{1}{r}\frac{du_\phi}{dr}+\frac{1}{r^2} \frac{d^2u_\phi}{d\theta^2}+\Delta k^2 u_\phi= \frac{\beta \epsilon_0   s\lambda g_{a\gamma \gamma} }{4 \pi  }\left(J_\nu \left(\frac{u}{a}r\right)^2+J_{\nu-1} \left(\frac{u}{a}r\right)J_{\nu+1} \left(\frac{u}{a}r\right)\right)\text{Sin}( 2(\nu \theta + \psi)).
\end{equation}

Using separation of variables, we can write our trial solution as $u_\phi(r,\theta)=R_\phi(r)\Theta (\theta)$ where we have defined the function $\Theta(\theta)=\text{Sin}( 2(\nu \theta + \psi))$. With this simplification, we obtain the following ordinary differential equation for $R_\phi(r)$ in the core of the fiber:

\begin{equation}\label{eq:axionDiffEq_core}
    \frac{d^2 R_\phi}{dr^2}+\frac{1}{r}\frac{dR_\phi}{dr}-\frac{4 \nu^2}{r^2}R_\phi +\Delta k^2 R_\phi= \gamma \left(J_\nu \left(\frac{u}{a}r\right)^2+J_{\nu-1} \left(\frac{u}{a}r\right)J_{\nu+1} \left(\frac{u}{a}r\right)\right).
\end{equation}

Using a similar procedure, the differential equation for the cladding part of the solution can be written as:

\begin{equation}\label{eq:axionDiffEq_clad}
    \frac{d^2 R_\phi}{dr^2}+\frac{1}{r}\frac{dR_\phi}{dr}-\frac{4 \nu^2}{r^2}R_\phi +\Delta k^2 R_\phi= \gamma \left(\frac{J_\nu(u)}{K_\nu(w)}\right)^2\left(K_\nu \left(\frac{w}{a}r\right)^2+K_{\nu-1} \left(\frac{w}{a}r\right)K_{\nu+1} \left(\frac{w}{a}r\right)\right).
\end{equation}

Here, we collect all the constants in the axion driving term into a single term and define:

\begin{equation}
    \frac{\beta \epsilon_0   s\lambda g_{a\gamma \gamma} }{4 \pi   }=\gamma
\end{equation}

\begin{figure}[htbp]
    \centering
    \includegraphics[width=\linewidth]{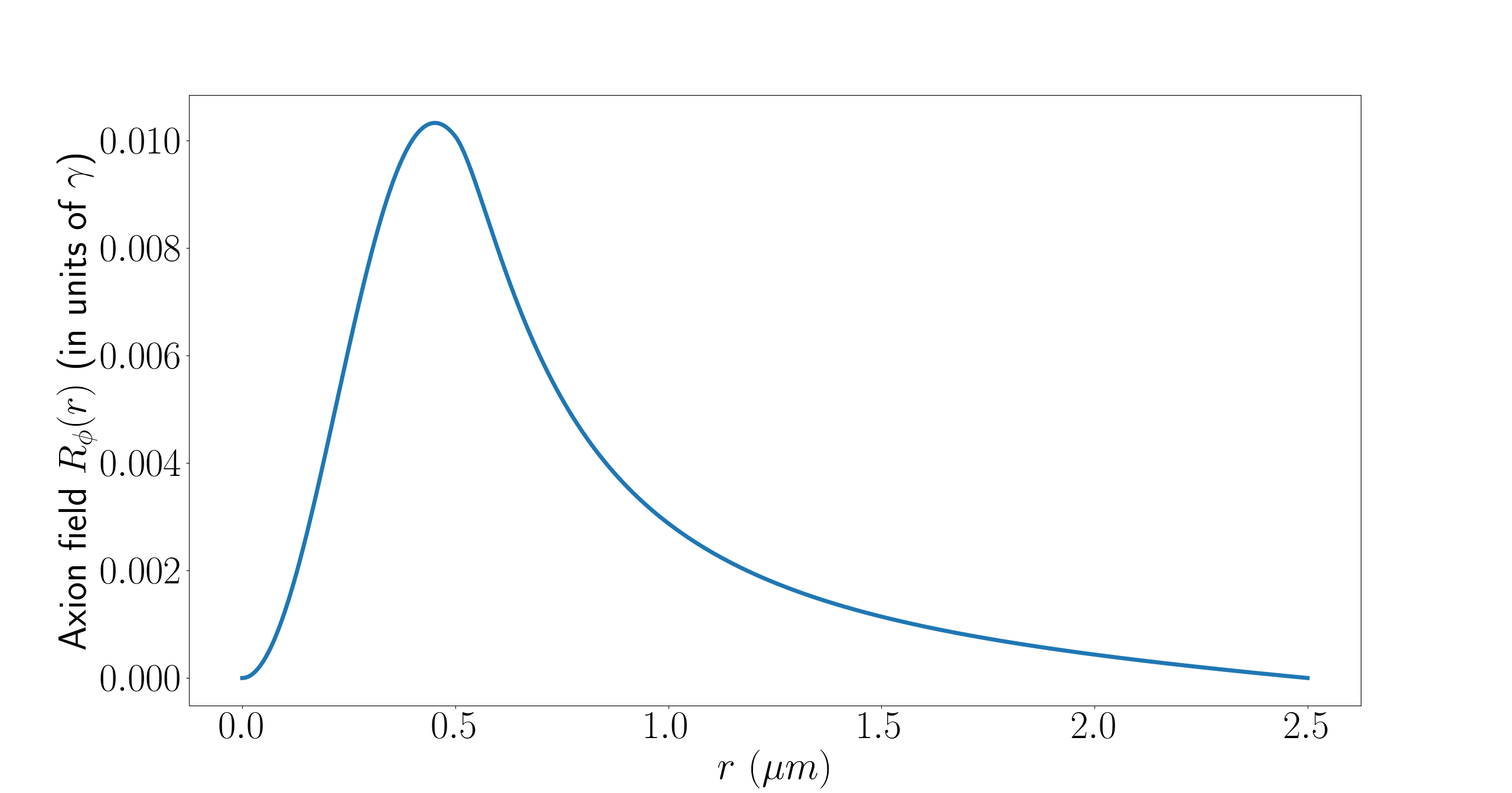}
    \caption{The Solution $R_\phi(r)$ with $\theta=\pi/4$ for the hypothetical axion field for the $HE_{11}$ mode with a propagation constant $\beta=13.60\times10^6m^{-1}$ for an air-clad fiber with radius a=0.5 $\mu m$ and a core index of refraction $n+1.46$ with a laser with wavelength $\lambda=633nm$ and $\Delta k^2=-65.05\times10^{6}m^{-2}$.}
    \label{fig:axionField-633}
\end{figure}

We numerically solve above Eq.~(\ref{eq:axionDiffEq_core}) and Eq.~(\ref{eq:axionDiffEq_clad}) using a fourth order Runge-Kutta algorithm for the non-transverse hybrid mode $HE_{11}$ that we have discussed above. We will present two examples for the confined axion mode profiles that are driven by the nontransverse electromagnetic modes. We first use a hybrid $HE_{11}$ mode with $\beta=13.60\times10^6m^{-1}$ of a laser at a wavelength $\lambda=633$~nm in an air clad silica fiber, with the index of refraction of $n_{core}=1.46$ and a core radius of $a=0.5\mu m$. For the axion mass $m=10^{-2}eV$, the value of $\Delta k^2$ is  $\Delta k ^2=-65.05\times10^{6}m^{-2}$. The solution of the axion differential equation with these constants can be seen on Fig.~\ref{fig:axionField-633}. The difference in the solutions as we scan the axion mass range from $m=10^{-2}eV$s to $m=10^{-6}eV$s is around $\gamma \times 10^{-8}$, which stays minimal, suggesting some solutions that are not very sensitive to the hypothetical axion mass. 

\begin{figure}[htbp]
    \centering
    \includegraphics[width=\linewidth]{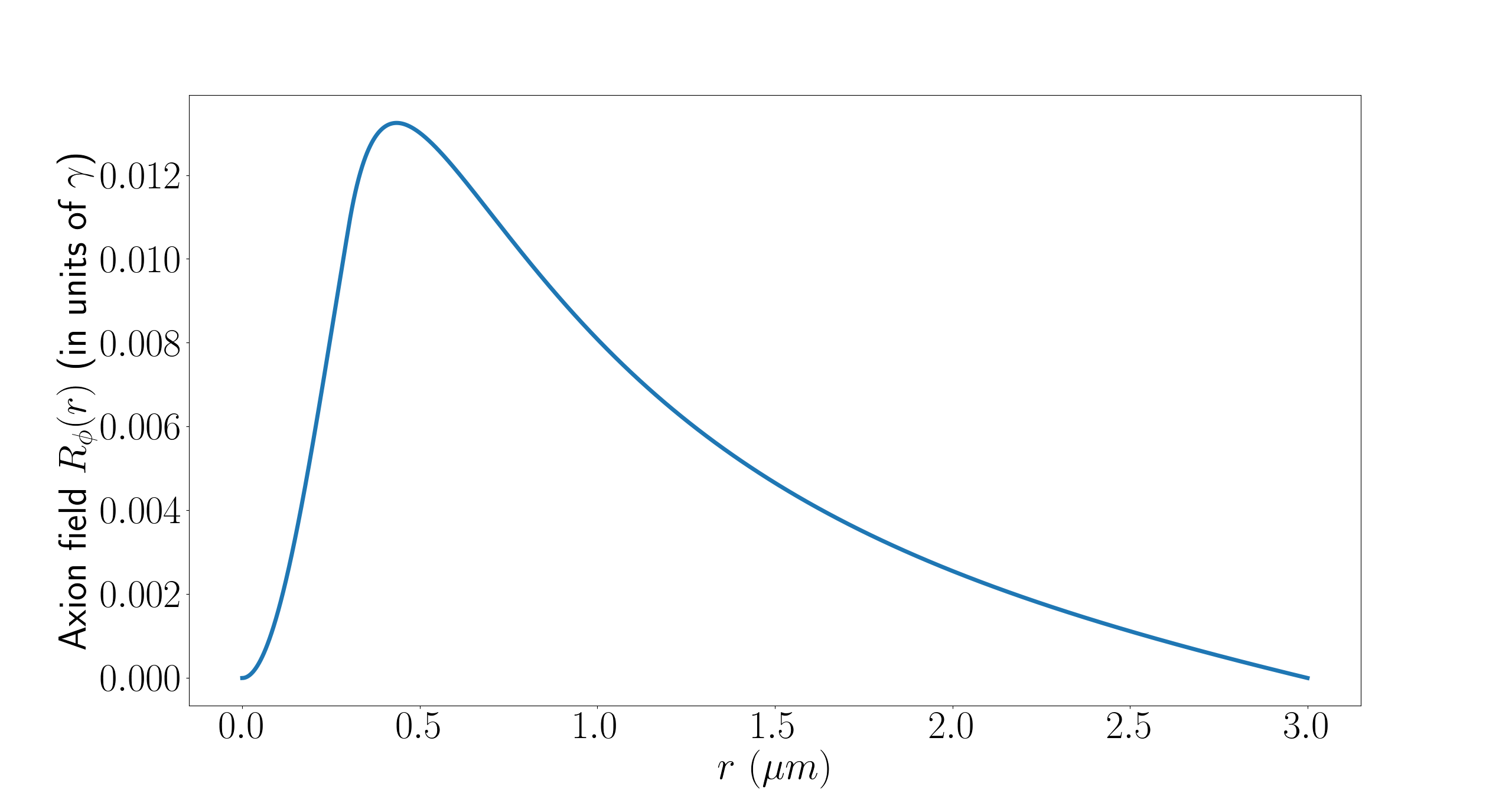}
    \caption{The Solution $R_\phi(r)$ with $\theta=\pi/4$ for the hypothetical axion field for the $HE_{11}$ mode with a propagation constant $\beta=4.16\times10^6m^{-1}$ for an air-clad fiber with radius a=0.3 $\mu m$ and a core index of refraction $n=1.46$ with a laser with wavelength $\lambda=1550nm$ and $\Delta k^2=-65.05\times10^{6}m^{-2}$.}
    \label{fig:axionField-1550}
\end{figure}

One can engineer the parameters of the fiber and the laser to obtain different spatial structures for the confined axion modes. Another example is shown in Fig.~\ref{fig:axionField-1550}. In this plot, the fiber structure has a radius of $a=0.3\mu m$, and an index of refraction of $n=1.46$, for a laser wavelength of $\lambda=1550nm$. The mode we choose for this case is the fundamental $HE_{11}$ mode with a a propagation constant of $\beta=4.16\times10^6m^{-1}$. Again, for these parameters we have $\Delta k^2=-65.05\times10^{6}m^{-2}$.

\begin{figure}[htbp]
    \centering
    \includegraphics[width=\linewidth]{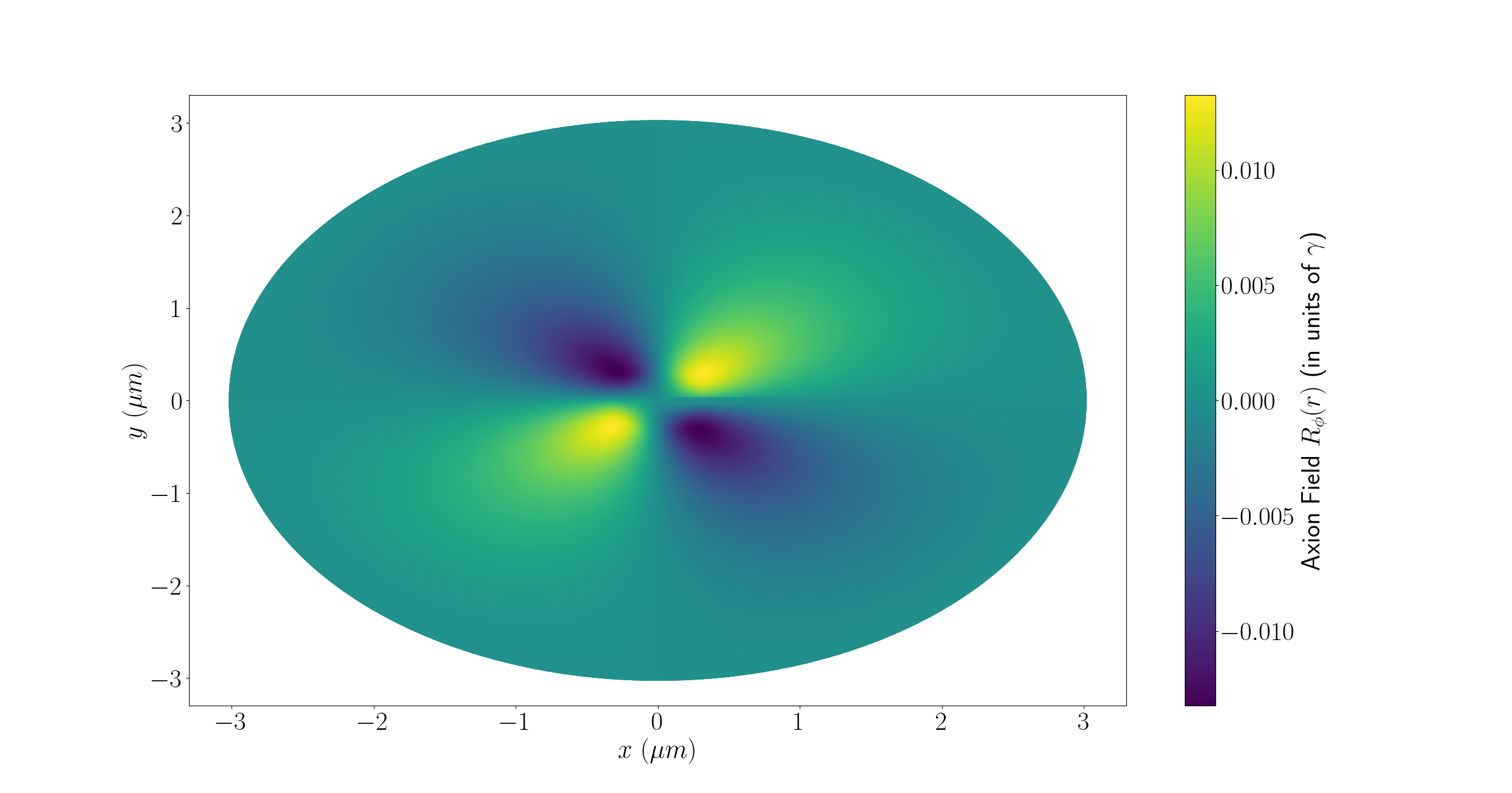}
    \caption{The 2D false color plot for the Solution for the hypothetical axion field $u_\phi(r, \theta)$ for the $HE_{11}$ mode ($\beta=4.16\times10^6m^{-1}$) for a fiber with radius $a=0.3\mu m$ and $n_{core}=1.46$ for a laser wavelength $\lambda=1550nm$ with $\Delta k^2=-0.89\times10^{12}m^{-2}$.}
    \label{fig:axionField-2D}
\end{figure}

To demonstrate the 2D behavior of the total field $u_\phi(r, \theta)=R_\phi(r)\Theta(\theta)$, we plot a 2D false color plot of the field $u_\phi(r, \theta)$ on Fig~\ref{fig:axionField-2D} for the parameters from the Fig.~\ref{fig:axionField-1550}. 

\section{A new type of optical traps for chiral atoms with magneto electric cross coupling}

\subsection{Chiral atoms with magneto electric cross coupling}

Over the last two decades, there has been a growing interest in materials that exhibit chirality through magneto-electric cross coupling. In traditional materials, the electric field of the light produces polarization of the medium while the magnetic field induces magnetization. In contrast, in materials with magneto-electric cross coupling, the electric field  can induce magnetization, while the magnetic field can polarize the medium. One application of these materials is to the studies of negative refractive index.  Materials that exhibit a negative index of refraction have been a subject of research because of their  interesting fundamental properties as well as applications that they demonstrate such as the possibility of constructing lenses whose performance can beat the diffraction limit~\cite{sikes_1, sikes, sikes_2, sikes_3, sikes_4, sikes_5, sikes_6, sikes_8, sikes_9, sikes_10, sikes_11, sikes_12}. One key challenge in constructing negative index materials is due to the weakness of the magnetic response in optical materials~\cite{sikes}. Chiral materials with magneto-electric cross coupling show considerable promise for overcoming this limitation due to additional contributions of the cross-coupling coefficients to the refractive index \cite{sikes_13, sikes_14}. Using such cross coupling scheme, one can achieve negative index of refraction without having a negative permeability~\cite{sikes,sikes_15,sikes_16}. 

In a material with magneto-electric cross coupling, the polarization, $P_p$, and the magnetization, $M_p$, for a probe beam with electric field $E_p$ and the magnetic field with $M_p$ is given by:

\begin{equation}
    \begin{split}
    P_p&=\epsilon_0 \chi_E E_p + \frac{\xi_{EB}}{c \mu_0 } B_p \quad , \\
    M_p&=\frac{\xi_{BE}}{c \mu_)} E_p + \frac{\xi_M}{\mu_0} B_p \quad .
    \end{split}   
\end{equation}

Here, the quantities $\chi_E$ and $\chi_M$ are the electric and the magnetic susceptibilities, and $\xi_{EB}$ and $\xi_{BE}$ are the complex cross-coupling (chirality) coefficients \cite{sikes, sikes_15, sikes_16}. 

\begin{figure}[htbp]
    \centering
    \includegraphics[width=\linewidth]{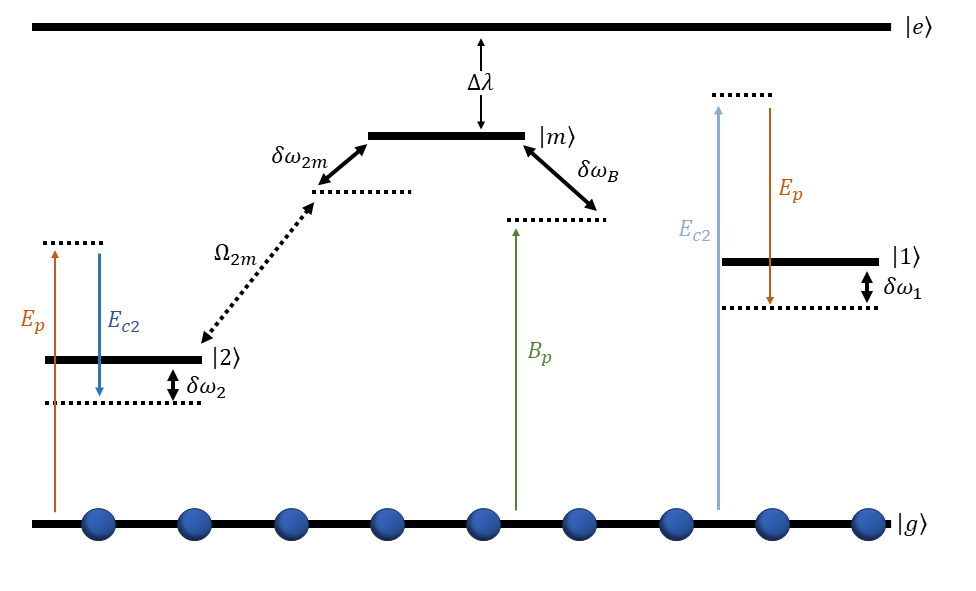}
    \caption{The energy level diagram of the suggested magneto-electric cross-coupling scheme~\cite{sikes}. The electric field $E_p$ and magnetic field $B_p$ are the EM components of the probe laser. $\ket{g}\rightarrow \ket{m}$ is a magnetic dipole transition induced by $B_p$. Two electric dipole Raman transitions are induced by the two strong control laser beams $E_{c1}$ and $E_{c2}$. These transitions can be detuned very far away from the excited state $\ket{e}$. Thus the system does not require $\ket{g}\rightarrow \ket{m}$ and $\ket{g}\rightarrow\ket{e}$ transitions to be close to the same frequency. The magneto-electric cross coupling(chirality) is induced by $\Omega_{2m}$. }
    \label{fig:chiral-diagram}
\end{figure}

In this section, we utilize the effect of magnetoelectric cross coupling-induced chirality for achieving a trap potential for atoms that scale with the term $\vec{E} \cdot \vec{B}$. For this purpose, we are going to focus on the specific five level scheme,  which was discussed in detail in Ref.~\cite{sikes}. As shown in Fig.~\ref{fig:chiral-diagram}, we consider an atomic system with a strong magnetic dipole transition with magnetic-moment $\mu_{gm}$ near the frequency of the probe laser beam. The system does not have a strong electric-dipole transition near the probe laser's frequency. Instead, the electric dipole response is obtained by using two-photon Raman transitions, that are detuned from the dipole-allowed excited state $\ket{e}$~\cite{sikes}. The two Raman transitions are introduced by using the probe laser and two intense control beams with electric fields $E_{c1}$ and $E_{c2}$, respectively. Because of the difference in the order of involvement of the probe laser in the Raman transitions, this method creates simultaneously an amplifying resonance and an absorbing resonance~\cite{sikes}. One can tune the strength and the position of these two Raman transitions by changing the frequencies and the intensities of the laser beams. The interference of the two resonances produces control of the index of refraction while having a small absorption.

In this scheme, the coherent coupling of the states $\ket{2}$ and $\ket{m}$ with a separate laser beam with Rabi frequency $\Omega_{2m}$ induces magneto-electric cross coupling. The states, $\ket{g}$, $\ket{1}$, $\ket{2}$ and $\ket{m}$ have the same parity, while being the opposite parity to the excited $\ket{e}$ state.  We define $\delta \omega_1=(\omega_{1}-\omega_g)-(\omega_{c1}-\omega_p)$ and $\delta \omega_2 = (\omega_2-\omega_g)-(\omega_p-\omega_{c2})$, which are the two-photon detunings. The detuning of the probe beam for the magnetic transition $\ket{g}\rightarrow\ket{m}$ can be defined as $\delta \omega_B = (\omega_m - \omega_g)-\omega_p$. Finally, the detuning of the magnetoelectric coupling laser from  the transition $\ket{2}\rightarrow\ket{m}$ is $\delta \omega_{2m}$. The system forms a closed loop and we therefore have $\delta \omega_B = \delta \omega_2 + \delta \omega_{2m}$. Following these definitions, one can derive the susceptibilities and the chirality coefficients as,

\begin{equation}
    \begin{split}
        \chi_E&=\frac{N}{\epsilon_0}\left[  \frac{|d_{ge}|^2}{\hbar \left(\Delta_p - i \Gamma_e/2\right)}+\frac{|d_{ge}|^2|d_{1e}|^2}{4 \hbar^3 \Delta_1^2 \left(\delta \omega_1 +i \gamma_1\right)}|E_{c1}|^2+\frac{|d_{ge}|^2|d_{2e}|^2}{4\hbar^3 \Delta_2^2 \left(\delta \omega_2 - \frac{|\Omega_{2m}|^2}{4(\delta \omega_B-i \gamma_m)}-i \gamma_2\right)}|E_{c2}|^2 \right]\\
        \xi_{EB}&=\mu_0 c N \frac{d_{ge} d^*_{2e} \mu^*_{gm}}{4\hbar^2 \Delta_2 (\delta \omega_B-i \gamma_m) \left(\delta \omega_2 - \frac{|\Omega_{2m}|^2}{4(\delta \omega_B - i \gamma_m)}-i \gamma_2\right)}\Omega_{2m}E_{c2}\\
    \end{split}
\end{equation}

with $N$ being the number of atoms per unit volume. Further details of this approach can be found from earlier publications of our group~\cite{sikes}.

\subsection{The chiral trap potential scaling with $\vec{E}\cdot \vec{B}$}

In this subsection, we introduce a trapping potential for chiral atoms that depends on the $U  \propto \vec{E} \cdot \vec{B}$, rather than the conventional dipole trap potential that scales with $U \propto |\vec{E}|^2$. We consider an atom with the above-described magneto-electric cross coupling scheme and, therefore, exhibiting chirality. The trapping potential for such an atom in the presence of the probe laser is:

\begin{equation}
   U_{trap}=\frac{1}{N}\langle \Vec{E}_p\cdot \vec{P}_p \rangle =  \frac{\epsilon_0 \chi_E}{N} | \vec{E}_p |^2 + \frac{\xi_{EB}}{N c \mu_0 } \langle \vec{E}_p \cdot \vec{B}_p \rangle,
\end{equation}

Using the results from the previous section and as we will discuss below, we have the freedom to tune the quantities $\chi_E$ and $\xi_{EB}$~\cite{sikes}. Therefore, one can find a condition where $\chi_E\ll\xi_{EB}$ and the trapping potential of Eq.~(18) is dominated by the second term. We note that for a trapping potential that is only proportional to $\langle \vec{E}_p \cdot \vec{B}_p \rangle$, because of the sign of the electromagnetic fields, there will be regions where the potential is attractive as well as repulsive. This behavior on its own is interesting and is different from the usual dipole potential $U \propto |\vec{E}|^2$ where depending on the sign of the detuning of the trapping laser from the excited state, the potential is either attractive or repulsive everywhere.

Having discussed the general scheme of the trapping, we proceed with a numerical example in an example atomic system. We take the excited state decay rate to be $\Gamma_e=2\pi \times 10$~MHz, and choose a probe laser with a wavelength $\lambda_{p}=600$~nm, two coupling lasers with wavelengths $\lambda_{c1}\approx\lambda_{c2}=750$~nm. We define the detunings as $\delta \omega_B = -\delta \omega_1 = \delta \omega_2$. We introduce a dephasing rate of  $\gamma_1=\gamma_2=\gamma_m=\gamma_c=2\pi \times 1$~MHz to the dipole-forbidden transition and take the Rabi frequency of the cross-coupling laser to be $\Omega_{2m}=i2\pi \times 2.12$~MHz. We use intensities of $I_{c1}=0.25$~MW/cm$^2$ and $I_{c2}=0.7$~MW/cm$^2$ for the two coupling lasers. We also note that for simplicity, the coupling lasers are taken uniform while the probe laser is assumed to be in the hybrid fiber mode analytically shown in Eq.~(\ref{eq:eq3}). We choose the optical power in the probe laser beam to be 1~W.  Finally, we calculate the electric and magnetic dipole matrix elements using the  Wigner-Weisskopf result

\begin{equation}
    d_{ij}=\sqrt{\frac{\pi \epsilon_0 \Gamma_e \hbar c^3}{\omega^3_p}},\qquad \mu_{ij}=\sqrt{\frac{\pi \epsilon_0 \Gamma_e \hbar c^5}{\omega^3_p 137^2}}.
\end{equation}

\begin{figure}[htbp]
    \centering
    \includegraphics[width=\linewidth]{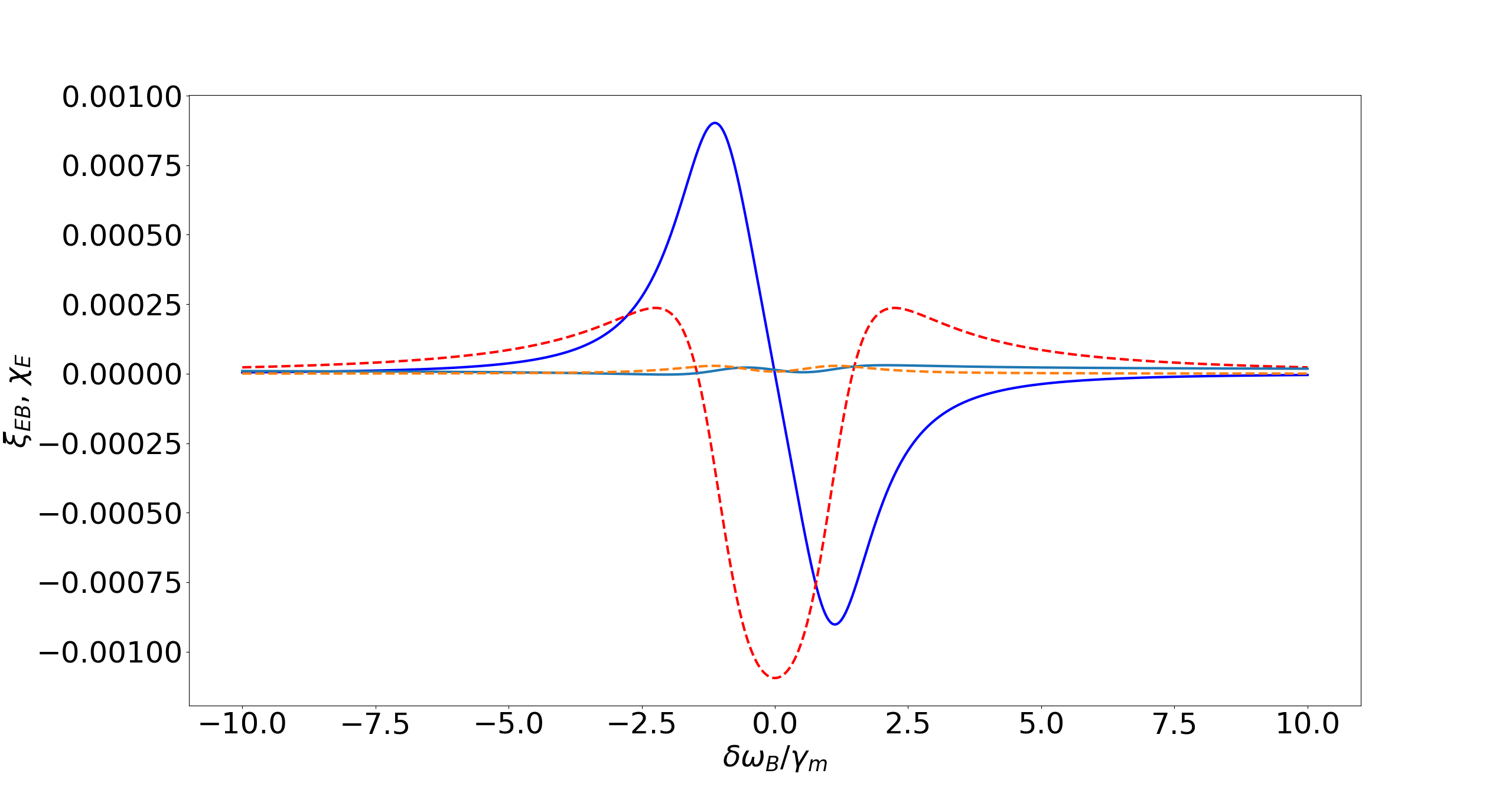}
    \caption{Susceptibility($\chi_E$) and Chirality coefficient ($\xi_{EB}$) calculated for a probe laser of $\lambda_{p}=600$~nm. The solid dark blue curve represents the real part of $\xi_{EB}$, solid light blue curve represents the real part of $\chi_{E}$, dashed red curve represents the imaginary part of the chirality coefficient $\xi_{EB}$ and dashed orange curve shows the imaginary part of the suscepteblity $\chi_{E}$.}
    \label{fig:chiE-xiEB}
\end{figure}

By using the parameters we presented, we plot the real part of the  susceptibility $Re\{\chi_E\}$ as solid light blue, and the imaginary part of the susceptibility as $Im\{\chi_E\}$ as dashed orange in Fig.~\ref{fig:chiE-xiEB}. In addition, the real part of the chirality constant $Re\{\xi_{EB}\}$ as solid blue and the imaginary part of the chirality $Im\{\xi_{EB}\}$ constant as dashed red again in Fig.~\ref{fig:chiE-xiEB} with respect to the  the detuning scaled with the decay rate $\delta \omega_B/\gamma_m$ for the probe laser.

\begin{figure}[htbp]
    \centering
    \includegraphics[width=\linewidth]{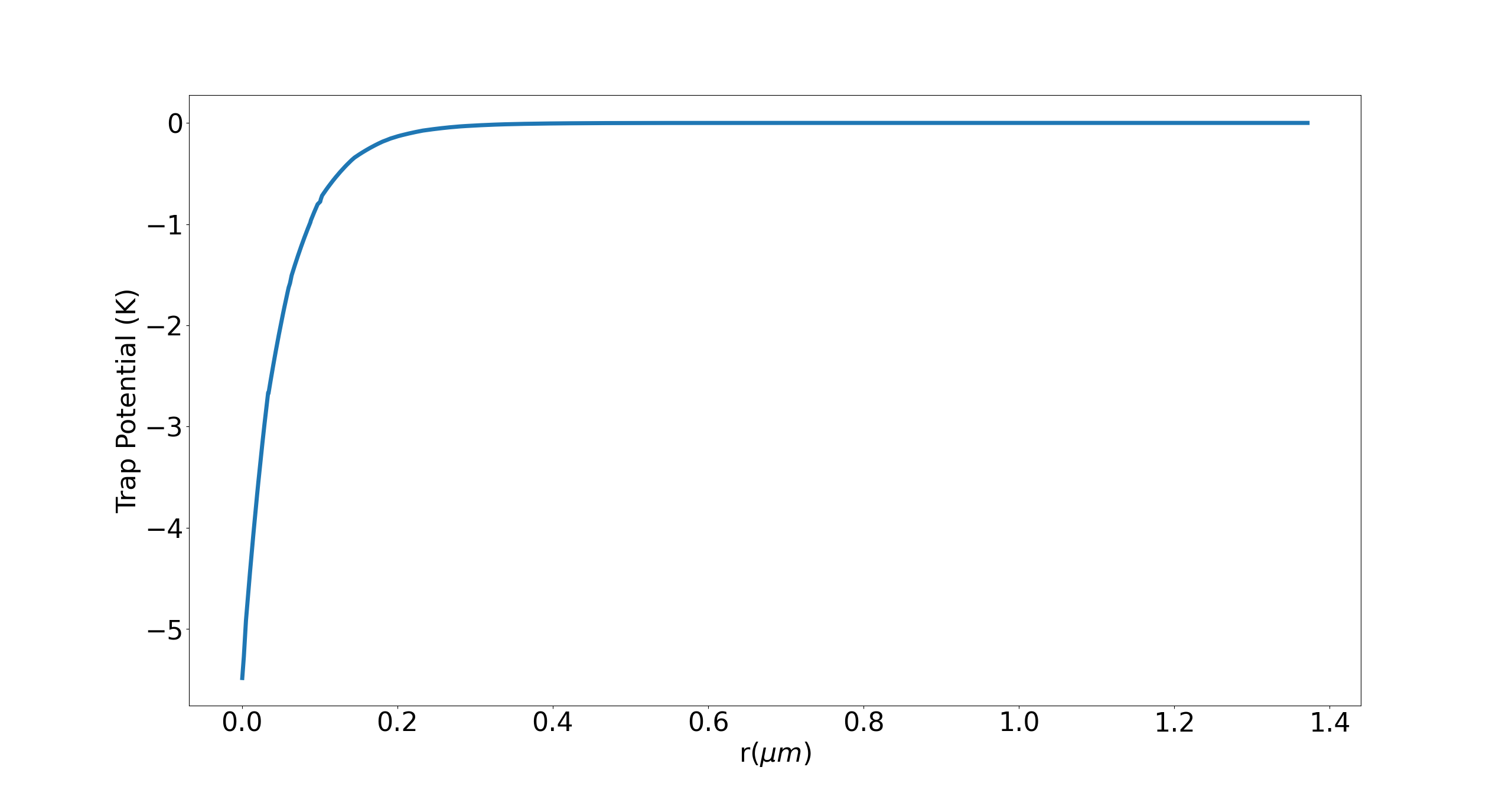}
    \caption{The calculation of the radial cross section of the chiral trap depth $U_{chiral}$ at the polar angle $\theta=\pi/4$ in Kelvins with the parameters  $\Gamma_e=2\pi \times 10$~MHz, $\lambda_{p}=600$~nm,  $\lambda_{c1}\approx\lambda_{c2}=750$~nm, $\delta \omega_B = -\delta \omega_1 = \delta \omega_2$, $\gamma_1=\gamma_2=\gamma_m=\gamma_c=2\pi \times 1$~MHz, $\Omega_{2m}=i2\pi \times 2.12$~MHz, $I_{c1}=0.25$~MW/cm$^2$ and $I_{c2}=0.7$~MW/cm$^2$.}
    \label{fig:trapDepth}
\end{figure}

From Fig.~\ref{fig:chiE-xiEB} we pick detuning $\delta \omega_{B}\approx -1.57 \gamma_m$ where the real parts of the susceptibility satisfy $\chi_E=0$.  Under this condition, we calculate the chiral trapping potential $U_{chiral}=\xi_{EB}/Nc\mu_0$ for $\lambda_p=600$~nm laser and take a linear cross section at the polar angle $\theta=\pi/4$ section of the fiber. We plot this linear potential radial cross section on Fig.~\ref{fig:trapDepth} in units of Kelvins. One can see that trap depths of 5K can reasonably be achieved. 

A common issue with attractive optical potentials originating from fiber modes is atoms being attracted and sticking on to the fiber's surface~\cite{atom-trap-5}. Therefore, one needs a balancing force to prevent the atoms from sticking onto the fiber's surface while maintaining a stable trap at a reasonable distance from the fiber surface. To obtain stable traps using the chiral potential approach, one approach would be to use two probe lasers, one being red-detuned and the other being blue detuned from the dipole-allowed excited level~\cite{atom-trap-5,atom-trap-8}. Using this approach, the attractive potential due to the red-detuned probe laser field very close to the surface of the fiber can be overcome by the repulsive potential of the blue-detuned trap laser. A detailed investigation of this approach will among our future investigations. 

\section{Conclusion}

In conclusion, we have presented an analytical and numerical study of electromagnetic modes in micro and nano fibers (MNFs) where the electric and magnetic fields of the propagating modes are not orthogonal to each other. We have also discussed applications of these modes in searching for hypothetical particles such as axions and in forming new type of optical traps for chiral atoms with magneto-electric cross coupling. 

One immediate future direction is to experimentally demonstrate optical traps where the confining potential is proportional to $\vec{E} \cdot \vec{B}$.  To demonstrate such traps, one approach would be to place the fiber structures that we have discussed above inside laser cooled atomic ensembles where magneto-electric cross coupling can be induced; for example lanthanides such as Terbium or Dysprosium. Another future direction is to integrate the above discussed fiber structures to proposed fiber and laser based axion searches and evaluate the bounds on axion-photon coupling that such experiments would place under realistic experimental conditions \cite{deniz_shay}.

\subsection{}{Funding}

This work was supported by the University of Wisconsin-Madison through Research Forward 1 award, and National Science
Foundation (NSF) Grant No. 2016136 for the QLCI
center Hybrid Quantum Architectures and Networks.

\subsection{Acknowledgments}
We thank Robby Rozite, Shay Inbar, and David Gold for many helpful discussions. 

\subsection{Disclosures}
The authors declare no conflicts of interest.

\subsection{Data Availability Statement}
The numerical data generated through COMSOL underlying the results presented in this paper are not publicly available at this time but may be obtained from the authors upon reasonable request.

\bigskip

\bibliography{ref}

\end{document}